\documentclass{article}
\usepackage{xcolor}
\usepackage{lineno}

\usepackage[english]{babel}
\usepackage[utf8]{inputenc}
\usepackage{amsmath}
\usepackage{amsfonts}
\usepackage{graphicx}
\usepackage[colorinlistoftodos]{todonotes}
\usepackage{blindtext}
\usepackage{amssymb}
\usepackage{caption,subcaption}
\usepackage{float}
\usepackage{hyperref}
\usepackage{multirow}
\usepackage[affil-it]{authblk}

\usepackage{amsthm}

\numberwithin{equation}{section}

\newtheorem{theorem}{Theorem}[section]

\newtheorem{lemma}{Lemma}[section]
\newtheorem{proposition}{Proposition}[section]
\newcommand{\tran}{\mathsf{T}}
\newcommand{\ml}{\mathcal{L}}
\newcommand{\mn}{\mathcal{N}}

\newcommand{\nid}{N_{i.}}
\newcommand{\nidb}{\bar{N}_{i.}}
\newcommand{\nldb}{\bar{N}_{l.}}

\newcommand{\ndj}{N_{.j}}
\newcommand{\ndjb}{\bar{N}_{.j}}
\newcommand{\ndsb}{\bar{N}_{.s}}
\newcommand{\ndkb}{\bar{N}_{.k}}

\newcommand{\dbern}{\mathrm{Bern}}
\newcommand{\dunif}{\mathrm{U}}

\newcommand{\zij}{Z_{ij}}
\newcommand{\pup}{\Upsilon}
\newcommand{\pij}{p_{ij}}
\newcommand{\stocleq}{\preccurlyeq}
\newcommand{\dbin}{\mathrm{Bin}}
\newcommand{\bone}{\mathbf{1}}

\newcommand{\azero}{a^{(0)}}
\newcommand{\aone}{a^{(1)}}
\newcommand{\atwo}{a^{(2)}}
\newcommand{\nipd}{N_{i'.}}
\newcommand{\sione}{s_{i}^{(1)}}
\newcommand{\sipone}{s_{i'}^{(1)}}

\newcommand{\slone}{s_{l}^{(1)}}

\newcommand{\ssone}{\sigma_1^{2}}

\newcommand{\sstwo}{\sigma_2^{2}}

\newcommand{\sse}{\sigma_E^{2}}
\newcommand{\sjtwo}{s^{(2)}_{j}}
\newcommand{\sjptwo}{s^{(2)}_{j'}}
\newcommand{\wione}{w_{i}^{(1)}}
\newcommand{\wjtwo}{w_{j}^{(2)}}
\newcommand{\ndjp}{N_{.j'}}

\newcommand{\simind}{\stackrel{\mathrm{ind}}{\sim}}
\newcommand{\wone}{\boldsymbol{w}^{(1)}}
\newcommand{\wtwo}{\boldsymbol{w}^{(2)}}
\newcommand{\baone}{\boldsymbol{a}^{(1)}}
\newcommand{\batwo}{\boldsymbol{a}^{(2)}}

\theoremstyle{remark}
\newtheorem*{remark}{Remark}

\theoremstyle{definition}

\title{Convergence rate of a collapsed Gibbs sampler for crossed random effects models}
\author{Swarnadip Ghosh$^*$  and Chenyang Zhong$^*$}
\affil{Stanford University}

\date{October, 2021}

\begin{document}
\maketitle
\def\thefootnote{*}\footnotetext{These authors contributed equally to this work}

\begin{abstract}
In this paper, we analyze the convergence rate of a collapsed Gibbs sampler for crossed random effects models. Our results apply to a substantially larger range of models than previous works, including models that incorporate missingness mechanism and unbalanced level data. The theoretical tools involved in our analysis include a connection between relaxation time and autoregression matrix, concentration inequalities, and random matrix theory. 
\end{abstract}

\section{Introduction}\label{Sect.1}



In this paper, we study the convergence rate of a collapsed Gibbs sampler for crossed random effects models. In applications of crossed random effects models (such as recommender systems), the missing data phenomenon can be quite common. Our analysis applies to models with missingness mechanism (see Sect.~\ref{Sect.1.2} for a detailed description), which is a new feature in the Bayesian approach to this problem. Another feature of our analysis is that it applies to data with unbalanced levels, which substantially relaxes the balancedness condition required for the analysis in \cite{papaspiliopoulos2020scalable}.




\subsection{Crossed random effects models}\label{Sect.1.1}
Crossed random effects models are regression models that relate a response variable to several categorical input variables. They are widely used in modeling enormous data sets coming from electronic commerce (such as recommender systems) and appear in literature under different names like cross-classified data, variance component models (\cite{gelman2005analysis,sear:case:mccu:1992}). 

In this paper, we consider the following crossed random effects model with two factors:
\begin{equation}
    Y_{ij}=a^{(0)}+a^{(1)}_i+a^{(2)}_j+e_{ij}, \quad 1\leq i\leq R, \quad 1\leq j\leq C.
\end{equation}
Here, $Y_{ij}$ is the response variable, $a^{(0)}$ is the global mean, $a^{(1)}_i$ and $a^{(2)}_j$ represent the two random effects, and $e_{ij}$ is the error. The first random effect has $R$ levels, and the second random effect has $C$ levels. We assume that $a^{(1)}_i\sim \mn(0,\sigma_1^2)$ for each $1\leq i\leq R$, $a^{(2)}_j\sim \mn(0,\sigma_2^2)$ for each $1\leq j\leq C$, and $e_{ij}\sim \mn(0,\sigma_E^2)$ for every $1\leq i\leq R,1\leq j\leq C$, where $\sigma_1,\sigma_2,\sigma_E>0$ (these random variables are assumed to be all independent). We denote the corresponding precision parameters by $\tau_1, \tau_2$ and $\tau_E$. We also assume a flat prior $p(a^{(0)})\propto 1$ for the global mean.

In our analysis, we assume that the observation $Y_{ij}$ for some levels $(i,j)$ (with $1\leq i\leq R$, $1\leq j\leq C$) can be missing. The missingness pattern is described by a $\{0,1\}$-valued matrix $Z$, which we describe in detail in Sect.~\ref{Sect.1.2}.


For recommender system applications, we let $i$ denote the $i$th customer for $1\leq i\leq R$, let $j$ denote the $j$th product for $1\leq j\leq C$, and let $Y_{ij}$ be the rating of the $i$th customer for the $j$th product. Then $a^{(1)}_i$ is the random effect from the $i$th customer, and $a^{(2)}_j$ is the random effect from the $j$th product.

In this paper, we are interested in scalable inference methods for crossed random effects models. Here, scalability means that the time complexity of the inference method should be at most linear in the number of observations and the number of parameters. This has been a great challenge for large data sets. For example, Gao and Owen \cite{gao2016estimation} showed that evaluating the likelihood of the model once already has complexity at least of order $N^{\frac{3}{2}}$ when there are $N$ observations. The lmer function in R package lme4 \cite{lme4} has a cost that grows like $N^{3/2}$ and Bates et al. \cite{lme4} removed the MCMC option from the package
because it was considered unreliable. As for MCMC algorithms, Gao and Owen \cite{gao2017efficient} showed that the Gibbs sampler takes order $N^{\frac{1}{2}}$ steps to mix, leading to order $N^{\frac{3}{2}}$ complexity (as the time cost for each iteration is of order $N$).

Liu \cite{liu1994collapsed} introduced collapsing in Monte Carlo computations. Recently, Papaspiliopoulos et al. \cite{papaspiliopoulos2020scalable} proposed a collapsed Gibbs sampler for crossed random effects models. They showed that under ``balanced levels'' condition the collapsed Gibbs sampler mixes in a constant number of steps in certain settings, and is therefore scalable. In the recommender system example, ``balanced levels'' condition reduces to each customer rating equal number of products and each item being rated by equal number of customers. We will review the collapsed Gibbs sampler and the balanced levels condition in Sect.~\ref{Sect.1.3}. Further discussions on the application of collapsed Gibbs sampling to Bayesian hierarchical models are in \cite{papaspiliopoulos2021scalable}. A different approach has been undertaken by Ghosh et al. \cite{ghosh2020backfitting}. They proposed an iterative algorithm based on backfitting and showed that in some regimes with unbalanced levels, the algorithm is scalable. Later they extended the algorithm to a generalized linear mixed model for logistic regression \cite{ghosh2021scalable}. They proposed \emph{clubbed backfitting}, a two step iterative procedure. It ``clubs'' the fixed effect and one of the random effects in each of the two steps.

The balanced levels condition is rarely satisfied in real applications such as recommender systems, unless we have a designed experiment for the data. Therefore, it is important to look for scalable inference methods that work beyond the balanced levels condition. In this paper, we show that the collapsed Gibbs sampler is scalable in many natural scenarios beyond the balanced levels condition. We relax the balanced levels condition to balanced in expectation, then
approximately balanced in expectation with some degree of inhomogeneity, then further relax it to arbitrary inhomogeneity assuming ``almost balancedness''. This motivates the study of this paper. As our approach, we analyse the $L_2$ norm of a certain autoregression matrix, which widens the range of scenarios where our result applies compared to the backfitting-based algorithm in \cite{ghosh2020backfitting} in a frequentist paradigm. Moreover, when the row sums and column sums are ``almost balanced'' in expectation (see Theorem \ref{Thm3} below for details), our proof strategy allows us to incorporate an \emph{arbitrary} inhomogeneity level (measured by $\Upsilon$ in Theorem \ref{Thm3}) for missingness pattern, which significantly improves upon the constraint in \cite{ghosh2020backfitting} (where it is required that $\Upsilon\leq 1.27$). The concrete results will be presented in Sect.~\ref{Sect.1.4}.


\subsection{Missingness mechanism}\label{Sect.1.2}

In this subsection, we describe the missingness mechanism of the model. The mechanism is similar to that of \cite{ghosh2020backfitting}, but is new in the Bayesian setting.

We let $Z_{ij}\in \{0,1\}$ for every $1\leq i\leq R$ and $1\leq j\leq C$, with $Z_{ij}=1$ when $Y_{ij}$ is observed and $Z_{ij}=0$ when $Y_{ij}$ is not observed. Now the $R\times C$ observation matrix $Z\in \{0,1\}^{R \times  C}$ formed by the elements $Z_{ij}$ has $N_{i\cdot}=\sum_{j=1}^C Z_{ij}$ observations in the $i$th row, and $N_{\cdot j}=\sum_{i=1}^R Z_{ij}$ observations in the $j$th column. The total number of observations is denoted by $N=\sum_{i=1}^R\sum_{j=1}^C Z_{ij}$.

In the following, we view $Z_{ij}$ as being non-random (i.e., we condition on the actual observations) when we perform collapsed Gibbs sampling. When studying convergence properties of the collapsed Gibbs sampler, we treat $Z_{ij}$ as independent random variables, with $Z_{ij}$ following the Bernoulli distribution with success probability $p_{ij}$ for every $1\leq i\leq R, 1\leq j\leq C$.

We mention by passing that our missingness mechanism does not incorporate informative missing, which can appear in real applications of crossed random effects models. However, as the analysis is already challenging under the current missingness mechanism, we leave the handling of informative missing to future works. 



\subsection{Collapsed Gibbs sampling}\label{Sect.1.3}

In \cite{papaspiliopoulos2020scalable}, Papaspiliopoulos et al. proposed a collapsed Gibbs sampler for sampling from the posterior of the crossed random effects model. They showed that under the balanced levels condition, the sampler is scalable. Here, a data set has balanced levels if for each factor, the number of observations for each level of the factor is the same (the number can vary with the factor). Via the notations from Sect.~\ref{Sect.1.2}, this condition can be expressed as $N_{i\cdot}=\frac{N}{R}$ for every $1\leq i\leq R$, and $N_{\cdot j}=\frac{N}{C}$ for every $1\leq j\leq C$.

Below we introduce the details of the collapsed Gibbs sampler, following \cite{papaspiliopoulos2020scalable}. We denote by $\baone=(a^{(1)}_1,\cdots,a^{(1)}_R)$, $\batwo=(a^{(2)}_1,\cdots,a^{(2)}_C)$. We also denote by $y$ the observations. The posterior distribution of $(\baone,\batwo)$ is denoted by $\mathcal{L}(\baone,\batwo\mid y)$. 

Each step of the collapsed Gibbs sampler involves integrating out the global mean $a^{(0)}$ and then sampling in blocks the levels of $\baone$ and $\batwo$. In practice, for the first block, we sample first from $\mathcal{L}(a^{(0)}|y,\batwo)$ and then from $\mathcal{L}(a^{(1)}_i|\cdot)$ for $i=1,\cdots,R$; for the second block, we sample first from $\mathcal{L}(a^{(0)}|y,\baone)$ and then from $\mathcal{L}(a^{(2)}_j|\cdot)$ for $j=1,\cdots,C$. 

We list the involved conditional distributions below, following \cite{papaspiliopoulos2020scalable}.

\begin{proposition}[\cite{papaspiliopoulos2020scalable}]
Let $\tilde{y}^{(k)}_j$ be the weighted average of all observations whose level on factor $k$ is $j$ for $k\in \{1,2\}$. For any $i=1,\cdots,R$, 
\begin{equation}
    \mathcal{L}(a^{(1)}_i|\cdot)=\mathcal{N}\{\frac{N_{i\cdot}\sigma_1^2}{N_{i\cdot}\sigma_1^2+\sigma_E^2}(\tilde{y}^{(1)}_i-a^{(0)}-\frac{\sum_{j=1}^C a^{(2)}_j Z_{i,j}}{N_{i\cdot}}),\frac{\sigma_E^2\sigma_1^2}{N_{i\cdot}\sigma_1^2+\sigma_E^2}\};
\end{equation}
for any $j=1,\cdots,C$,
\begin{equation}
    \mathcal{L}(a^{(2)}_j|\cdot)=\mathcal{N}\{\frac{N_{\cdot j}\sigma_2^2}{N_{\cdot j}\sigma_2^2+\sigma_E^2}(\tilde{y}^{(2)}_j-a^{(0)}-\frac{\sum_{i=1}^R a^{(1)}_i Z_{i,j}}{N_{\cdot j}}),\frac{\sigma_E^2\sigma_2^2}{N_{\cdot  j}\sigma_2^2+\sigma_E^2}\}.
\end{equation}

For any $i=1,\cdots,R$, we let $s^{(1)}_i=\frac{N_{i\cdot}\sigma_1^2 }{N_{i\cdot}\sigma_1^2 +\sigma_E^2}$; for any $j=1,\cdots,C$, we let $s^{(2)}_j=\frac{N_{\cdot j}\sigma_2^2}{N_{\cdot j}\sigma_2^2+\sigma_E^2}$. Then we have
\begin{equation}
    \mathcal{L}(a^{(0)}|y,\batwo)=\mathcal{N}(\frac{1}{\sum_{i=1}^R s^{(1)}_i}\sum_{i=1}^R s^{(1)}_i(\tilde{y}^{(1)}_i-\frac{\sum_{j=1}^C a^{(2)}_j Z_{i,j}}{ N_{i\cdot} }),\frac{\sigma_1^2}{\sum_{i=1}^R s^{(1)}_i}),
\end{equation}
\begin{equation}
    \mathcal{L}(a^{(0)}|y,\baone)=\mathcal{N}(\frac{1}{\sum_{j=1}^C s^{(2)}_j}\sum_{j=1}^C s^{(2)}_j(\tilde{y}^{(2)}_j-\frac{\sum_{i=1}^R a^{(1)}_i Z_{i,j}}{ N_{\cdot j} }),\frac{\sigma_2^2}{\sum_{j=1}^C s^{(2)}_j}).
\end{equation}
\end{proposition}

\subsection{Main results}\label{Sect.1.4}

In this section, we describe the main results of this paper. We measure the convergence rate of the collapsed Gibbs sampler by the relaxation time, denoted by $t_{rel}$. We recall that the relaxation time of a Markov chain is defined as the reciprocal of its spectral gap, where the spectral gap is the difference between $1$ and the geometric rate of convergence of the Markov chain. The readers are referred to \cite{papaspiliopoulos2020scalable,roberts1997updating} for further details. 

We show that in three regimes, the collapsed Gibbs sampler mixes in a \emph{constant} number of steps (independent of the problem size) in terms of relaxation time with high probability (with respect to the randomness in the observation matrix $Z$). Therefore, the collapsed Gibbs sampler is scalable for these regimes.

\begin{figure}[H]
  \centering
   \includegraphics[width=.8\hsize]{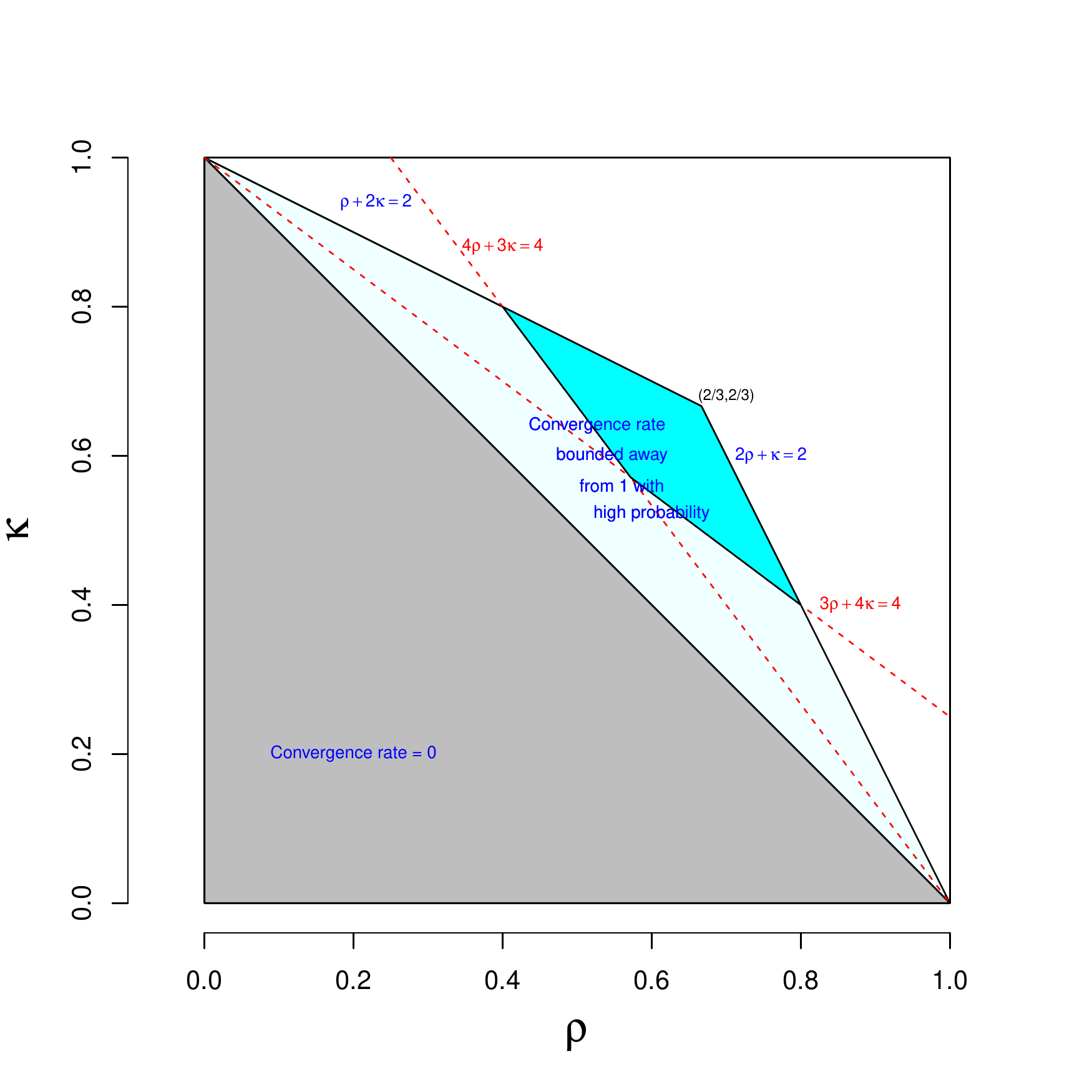}
\caption{\label{fig:domainofinterest}
The area shaded with light blue and cyan is the domain of interest for
Theorems \ref{Thm1}-\ref{Thm3}. Ghosh et al. \cite{ghosh2020backfitting} proved convergence for the backfitting procedure for $(\rho,\kappa)$ in the light blue region.}
\end{figure}

In our analysis, we assume that $Z_{ij}$ follows independent Bernoulli distribution with success probability $p_{ij}$ for each $1\leq i\leq R, 1\leq j\leq C$. Our problem size will be indexed by $S$, and we let $S\rightarrow\infty$ in the asymptotic analysis. The total sum $N$ of the matrix $Z$ satisfies $\mathbb{E}[N]\asymp S$. We also assume the scaling $R=\lceil S^{\rho} \rceil,C=\lceil S^{\kappa}\rceil$ for $\rho,\kappa\in (0,1)$.  We are interested in regimes where the observations are sparse, that is, $\rho+\kappa>1$ (so that $S\ll RC$). Our analysis holds for $\rho,\kappa\in (0,1)$ such that 
\begin{eqnarray}\label{regime}
 \rho+\frac{1}{2}\kappa<1, \quad
 \kappa+\frac{1}{2}\rho<1.
\end{eqnarray}
This is wider than the regime covered by the analysis of \cite{ghosh2020backfitting}. The range of $\rho,\kappa$ covered by our analysis is indicated in Figure \ref{fig:domainofinterest}. For the theoretical analysis we assume the precision parameters to be known. Empirically we observe a similar phenomenon when they are unknown and we sample precision parameters $\tau$ from the conditional distribution $\mathcal{L}(\tau \mid \baone,\batwo)$ and update $(\baone, \batwo)$ with the Gibbs sampler and its collapsed version, respectively.


The first result assumes that $Z_{ij}$ are i.i.d. This corresponds to the situation where the missingness pattern is homogeneous across different cells $(i,j)$ for $1\leq i\leq R,1\leq j\leq C$. This missingness mechanism is known as missing completely at random (MCAR) \cite{little2019statistical}.

\begin{theorem}\label{Thm1}
Suppose that $Z_{ij}  \sim \dbern(\frac{S}{RC})$ for every $1\leq i\leq R$ and $1\leq j\leq C$. Assume that $\rho$ and $\kappa$ satisfy condition \eqref{regime}, and that $\sigma_1^2,\sigma_2^2,\sigma_E^2$ are independent of $S$. 

Then there exists a positive constant $\xi$ (independent of $S$), such that
\begin{equation*}
    \lim_{S\rightarrow\infty}\mathbb{P}(t_{rel}\leq \xi)=1.
\end{equation*}
\end{theorem}

\begin{proof}
It follows from Proposition \ref{prop:collapsedgibbsmatrix} and Theorem~\ref{theo:theo1} below.
\end{proof}

The second result assumes that $\frac{S}{RC}\leq p_{ij} \leq \frac{\Upsilon S}{RC}$ with $\Upsilon\in [1,1.52]$. In this regime, we allow inhomogeneous missingness pattern across different cells $(i,j)$ for $1\leq i\leq R,1\leq j\leq C$. This missingness mechanism corresponds to missing at random (MAR) \cite{little2019statistical}.

\begin{theorem}\label{Thm2}
Assume that $\rho$ and $\kappa$ satisfy condition \eqref{regime}, and that $\sigma_1^2,\sigma_2^2,\sigma_E^2$ are independent of $S$. Also assume that there exists a constant $\Upsilon\in [1,1.52]$, such that for every $1\leq i\leq R$ and $1\leq j\leq C$,
\begin{equation*}
    \frac{S}{RC}\leq p_{ij}\leq \frac{\Upsilon S}{RC}.
\end{equation*}
Then there exists a positive constant $\xi$ (independent of $S$), such that
\begin{equation*}
    \lim_{S\rightarrow\infty}\mathbb{P}(t_{rel}\leq \xi)=1.
\end{equation*}
\end{theorem}
\begin{proof}
It follows from Proposition \ref{prop:collapsedgibbsmatrix} and Theorem~\ref{theo:theo2} below.
\end{proof}
\begin{remark}
Theorem \ref{Thm1} is a special case of Theorem \ref{Thm2} with $\Upsilon=1$.
\end{remark}



The third result assumes that the row sums and column sums of the matrix $Z$ are ``almost balanced'' in expectation (given by the conditions (\ref{balance1})-(\ref{balance2}) below). In the recommender system application, this corresponds to the case when every customer\slash product has similar behavior in expectation. With this almost balancedness condition, we can allow for an arbitrary inhomogeneity parameter $\Upsilon>1$ (with $\frac{S}{\Upsilon RC}\leq p_{ij}\leq \frac{\Upsilon S}{RC}$). This missingness mechanism also corresponds to missing at random (MAR) \cite{little2019statistical}.

\begin{theorem}\label{Thm3}
Assume that $\rho$ and $\kappa$ satisfy condition \eqref{regime}, and that $\sigma_1^2,\sigma_2^2,\sigma_E^2$ are independent of $S$. Also assume that there exist a series of numbers $\epsilon(S)>0$ with $\lim\limits_{S \rightarrow  \infty}\epsilon(S)=0$, such that for every $1\leq i\leq R$ and $1\leq j\leq C$,
\begin{equation}\label{balance1}
  \frac{S}{C}(1-\epsilon(S)) \leq  \sum_{i=1}^R p_{ij} \leq\frac{S}{C}(1+\epsilon(S)),
\end{equation}
\begin{equation}\label{balance2}
  \frac{S}{R}(1-\epsilon(S)) \leq  \sum_{j=1}^C p_{ij} \leq \frac{S}{R}(1+\epsilon(S))).
\end{equation}
Further assume that there exists a constant $\Upsilon\geq 1$, independent of $S$, such that for every $1\leq i\leq R$ and $1\leq j\leq C$,
\begin{equation*}
    \frac{S}{\Upsilon RC}\leq p_{ij}\leq \frac{\Upsilon S}{RC}.
\end{equation*}

Then there exists a positive constant $\xi$ (independent of $S$), such that
\begin{equation*}
    \lim_{S\rightarrow\infty}\mathbb{P}(t_{rel}\leq \xi)=1.
\end{equation*}
\end{theorem}

\begin{proof}
It follows from Proposition \ref{prop:collapsedgibbsmatrix} and Theorem~\ref{theo:theo3} below.
\end{proof}

The proof strategy of our main results is substantially different from that used in \cite{papaspiliopoulos2020scalable}. Specifically, the theoretical analysis of \cite{papaspiliopoulos2020scalable} is based on multigrid decomposition, which crucially depends on the balanced levels condition. In our proof, instead, we utilize a connection between the mixing time of the collapsed Gibbs sampler and the autoregression matrix of the chain without performing a multigrid decomposition (see Sect.~\ref{Sect.2} for details).
The analysis of the autoregression matrix, as given in Sect.~\ref{Sect.3}, involves many further tools, including concentration inequalities and results from random matrix theory. Equipped with these tools, our analysis provides sufficient conditions for convergence within a finite number of steps in the three settings covered by Theorems \ref{Thm1}--\ref{Thm3}. In comparison, the convergence results of \cite{papaspiliopoulos2020scalable} depend on the rate of convergence of an auxiliary Markov chain (denoted by $\rho_{aux}$ in the paper), for which not much is known in many practical settings.





Now we describe the main contents of the following sections. Sect.~\ref{Sect.2} introduces our main strategy for studying the relaxation time of the collapsed Gibbs sampler with unbalanced levels. Applying this strategy to three concrete scenarios, Sect.~\ref{Sect.3} provides relaxation time upper bounds in such settings, which show that the collapsed Gibbs sampler is scalable in these settings. Then through numerical simulations, we show in Sect.~\ref{Sect.4} the scalability of the collapsed Gibbs sampler in various settings, even beyond those covered by our theoretical analysis. In Sect.~\ref{Sect.5}, we further illustrate the scalability of the collapsed Gibbs sampler on real data from Stitch Fix. Finally, in Sect.~\ref{Sect.6}, we present some conclusions and discussions.

\section{Relaxation time and autoregression matrix}\label{Sect.2}
Suppose the chain $\left(\baone(t),\batwo(t)\right)$ is a two-component Gibbs sampler that alternates updates from
$\ml(\baone | y, \batwo)$ and $\ml(\batwo | y, \baone)$. Thus, $\batwo(t)$ is marginally a Markov chain and its rate
of convergence equals that of $\left(\baone(t),\batwo(t)\right)$ (see, for instance, \cite{robertsahoo2001}). Let $B_1$ and
$B_2$ be defined by $\mathbb{E}(\baone | \batwo, y) = B_1\batwo +b_1$ and $\mathbb{E}(\batwo | \baone, y) = B_2\baone +b_2$. Then $\batwo(t)$ is a Gaussian autoregressive process with autoregression
matrix $B_2B_1$. By \cite[Theorem 1]{roberts1997updating}, to study the relaxation time, it suffices to look at the autoregression matrix.

We consider a crossed random effects model with two factors (that is, $K = 2$) as in the Introduction. 
Let $\lambda_A = \frac{\sse}{\ssone}$ and $\lambda_B = \frac{\sse}{\sstwo}$. Then $\sione = \frac{\nid}{\nid + \lambda_A}$ and $\sjtwo = \frac{\ndj}{\ndj+\lambda_B}.$ Along with these, let us define the following two sets of weights, $\wione = \frac{\sione}{\sum_{i'}\sipone}$ and $\wjtwo = \frac{\sjtwo}{\sum_{j'}\sjptwo}$ and the two diagonal matrices $D_1 = \mathrm{diag}(\nid + \lambda_A)$, $D_2 = \mathrm{diag}(\ndj + \lambda_B).$ Throughout the paper, for any $n\in\{1,2,\cdots\}$, we denote by $I_n$ the identity matrix of size $n$ and $\bone_n$ the $n$-dimensional vector whose entries are all $1$. 

\begin{proposition}\label{prop:collapsedgibbsmatrix}
The relaxation time $t_{rel}$ of the collapsed Gibbs sampler is characterised by the spectral radius $\rho(M)$ of the doubly centered matrix $M:=(I_{C} - \wtwo \mathbf{1}_C^{\tran})M_0$ (with $M_0:=D_2^{-1}Z^{\tran}(I_{R} - \wone \mathbf{1}_R^{\tran})D_1^{-1}Z$) by 
\begin{equation}
    t_{rel}=\frac{1}{1-\rho(M)}.
\end{equation}
\end{proposition}

\begin{proof}
See Appendix~\ref{sec:proof:prop:collapsedgibbsmatrix}.
\end{proof}

\section{Analysis of the autoregression matrix}\label{Sect.3}

For the following analysis we would assume the following model. We let $\zij\sim\dbern(\pij)$ for every $1\leq i\leq R$ and $1\leq j\leq C$ independently with
\begin{align}\label{eq:defab}
\frac{1}{\pup'}\frac{S}{RC} \le \pij \le \pup\frac{S}{RC}
\quad\text{for}\quad 1\le\pup, \pup'<\infty, R = S^\rho, C = S^\kappa.
\end{align}
That is $1/\pup' \le \pij S^{\rho+\kappa-1}\le\pup$.
Letting $\pij$ depend on $i$ and $j$
allows the probability model to capture
stylistic preferences affecting the missingness
pattern in the ratings data. Note that we have ignored the issue of taking integer parts for $R$ and $C$, as it will not influence our conclusion.

For any matrix $A$, we denote by $\Vert A\Vert_2$ its spectral norm, and denote by $\rho(A)$ its spectral radius below. The meanings of variables like $\delta_1,\delta_2, \epsilon$ defined in proofs can vary across different proofs, but will remain the same within one proof.

For the three regimes discussed in Sect.~\ref{Sect.1.4}, the corresponding results for the autoregression matrix are given by Theorems \ref{theo:theo1}-\ref{theo:theo3} as follows. Theorems \ref{Thm1}-\ref{Thm3} follow from combining Theorems \ref{theo:theo1}-\ref{theo:theo3} and Proposition \ref{prop:collapsedgibbsmatrix}.

\begin{theorem}\label{theo:theo1}
Assume that $\rho$ and $\kappa$ satisfy condition \eqref{regime}, and that $\sigma_1^2,\sigma_2^2,\sigma_E^2$ are independent of $S$. Also assume that $\Upsilon'=\Upsilon=1.$

Then for any fixed $\delta>0$, we have
\begin{equation*}
    \lim_{S\rightarrow\infty}\mathbb{P}(\Vert M\Vert_2\leq \delta)=1.
\end{equation*}
\end{theorem}

\begin{proof}
See Section~\ref{Sect.3.2}.
\end{proof}

\begin{theorem}\label{theo:theo2}
Assume that $\rho$ and $\kappa$ satisfy condition \eqref{regime}, and that $\sigma_1^2,\sigma_2^2,\sigma_E^2$ are independent of $S$. Also assume that $\Upsilon'=1$ and $\Upsilon\in [1,1.52]$. Take
\begin{equation*}
    \phi(\Upsilon)=\frac{1}{\Upsilon^3}-(\Upsilon-1)^2.
\end{equation*}

Then for any fixed $\delta>0$, we have
\begin{equation*}
    \lim_{S\rightarrow\infty}\mathbb{P}\bigl(\rho(M)\leq 1-\phi(\Upsilon)+\delta\bigr)=1.
\end{equation*}
\end{theorem}
\begin{proof}
See Section~\ref{Sect.3.3}.
\end{proof}

\begin{remark}
We note that for $\Upsilon\in [1,1.52]$, $\phi(\Upsilon)\geq 0.0143>0$.
\end{remark}

\begin{theorem}\label{theo:theo3}
Assume that $\rho$ and $\kappa$ satisfy condition \eqref{regime}, $\Upsilon'=\Upsilon$, and $\sigma_1^2,\sigma_2^2,\sigma_E^2,\Upsilon$ are independent of $S$. Also assume that there exist a series of numbers $\epsilon(S)>0$ with $\lim\limits_{S  \rightarrow  \infty}\epsilon(S)=0$, such that for every $1\leq i\leq R$ and $1\leq j\leq C$,
\begin{equation*}
  \frac{S}{C}(1-\epsilon(S)) \leq  \sum_{i=1}^R p_{ij} \leq\frac{S}{C}(1+\epsilon(S)),
\end{equation*}
\begin{equation*}
  \frac{S}{R}(1-\epsilon(S)) \leq  \sum_{j=1}^C p_{ij} \leq \frac{S}{R}(1+\epsilon(S)).
\end{equation*}

Then there exists $\delta>0$ (independent of $S$), such that
\begin{equation*}
    \lim_{S\rightarrow\infty}\mathbb{P}(\rho(M)\leq  1-\delta)=1.
\end{equation*}
\end{theorem}
\begin{proof}
See Section~\ref{Sect.3.4}.
\end{proof}

The rest of this section is devoted to the proofs of Theorems \ref{theo:theo1}-\ref{theo:theo3}, in Sect.~\ref{Sect.3.2}-\ref{Sect.3.4}. Some preparatory results are presented in Sect.~\ref{Sect.3.1}. For any $n\in\{1,2,\cdots\}$, we denote by $[n]=\{1,2,\cdots,n\}$.

\subsection{Preparatory results}\label{Sect.3.1}

In this section, we present several preparatory results, which will be used in the proofs of Theorems \ref{theo:theo1}--\ref{theo:theo3}.

\begin{lemma}\label{lem:hoeff}
If $X\sim\dbin(n,p)$, then for any $t\ge0$,
\begin{align*}
\mathbb{P}( X\ge np+t ) &\le \exp( -2t^2/n ),\quad\text{and}\\
\mathbb{P}( X\le np-t ) &\le \exp( -2t^2/n ).
\end{align*}
\end{lemma}
\begin{proof}
This follows from Hoeffding's theorem.
\end{proof}

\begin{lemma}\label{lem:rowcolumn}
Under the model in~\eqref{eq:defab}, for any $\psi>0$,
\begin{align}
&\mathbb{P}\Bigl( (\frac{1}{\pup'}-\psi) S^{1-\rho}\le \min_i \sum_j \zij \le \max_i \sum_j \zij \le (\pup+\psi) S^{1-\rho}\Bigr)\nonumber\\
&\geq  1-2R\exp(-2S^{2-\kappa-2\rho}\psi^2).
\end{align}
Hence if $\rho+\frac{1}{2}\kappa<1$, then for any fixed $\psi>0$,
\begin{align}\label{eq:boundnid}
\lim_{S\to\infty}\mathbb{P}\Bigl( (\frac{1}{\pup'}-\psi) S^{1-\rho}\le \min_i \sum_j \zij \le \max_i \sum_j \zij \le (\pup+\psi) S^{1-\rho}\Bigr)=1.
\end{align}
Likewise, for any $\psi>0$,
\begin{align}
&\mathbb{P}\Bigl( (\frac{1}{\pup'}-\psi)S^{1-\kappa}\le \min_j \sum_i \zij \le \max_j \sum_i \zij \le (\pup+\psi) S^{1-\kappa}\Bigr)\nonumber\\
&\geq 1-2C \exp(-2S^{2-\rho-2\kappa}\psi^2).
\end{align}
Hence if $\frac{1}{2}\rho+\kappa<1$, then for any fixed $\psi>0$,
\begin{align}\label{eq:boundndj}
\lim_{S\to\infty}\mathbb{P}\Bigl( (\frac{1}{\pup'}-\psi)S^{1-\kappa}\le \min_j \sum_i \zij \le \max_j \sum_i \zij \le (\pup+\psi) S^{1-\kappa}\Bigr)=1.
\end{align}
\end{lemma}

\begin{proof}
See Appendix~\ref{sec:proof:lem:rowcolumn}.
\end{proof}

\begin{lemma}\label{lem:Zeigen}
$\|Z\|_2\leq\sqrt{\max_i\nid\max_j\ndj}$. Under the model in~\eqref{eq:defab} and the condition in ~\eqref{regime},
with high probability $\|Z\|_2\leq \frac{(\Upsilon+1)S}{\sqrt{RC}}.$
\end{lemma}
\begin{proof}
See Appendix~\ref{sec:proof:lem:eigenv}
\end{proof}

The following bound on spectral norms of random matrices was proved by Rafa{\l} Lata{\l}a \cite{Latala}.  

\begin{proposition}\label{prop:RMT}[\cite{Latala}, Theorem 2]
Suppose $X$ is a random matrix with independent mean zero entries $X_{ij}$. Then
\begin{equation}
   \mathbb{E}\Vert X\Vert_2 \leq K\left(\max _{i} \sqrt{\sum_{j} \mathbb{E} X_{i j}^{2}}+\max _{j} \sqrt{\sum_{i} \mathbb{E} X_{i j}^{2}}+\sqrt[4]{\sum_{i j} \mathbb{E} X_{i j}^{4}}\right),
\end{equation}
where $K>0$ is an absolute constant.
\end{proposition}

\begin{lemma}\label{lem:RMT}
Under the model in~\eqref{eq:defab}, 
\begin{equation}
    \mathbb{E}\Vert Z-\mathbb{E}(Z)\Vert_2\leq K\left(\sqrt{\frac{\Upsilon S}{R}}+\sqrt{\frac{\Upsilon S}{C}}+\sqrt[4]{\Upsilon S}\right),
\end{equation}
where $K>0$ is an absolute constant.
\end{lemma}
\begin{proof}
For any $1\leq i\leq R$ and $1\leq j\leq C$, 
\begin{equation}
    \mathbb{E}[(Z_{ij}-p_{ij})^2]=p_{ij}(1-p_{ij})\leq p_{ij}\leq \frac{\Upsilon S}{RC},
\end{equation}
\begin{equation}
    \mathbb{E}[(Z_{ij}-p_{ij})^4]=p_{ij}(1-p_{ij})((1-p_{ij})^3+p_{ij}^3)\leq p_{ij}\leq \frac{\Upsilon S}{RC}.
\end{equation}
As $\{Z_{ij}-p_{ij}\}_{1\leq i\leq R, 1\leq j\leq C}$ are independent with zero mean, we can apply Proposition \ref{prop:RMT} and get
\begin{align}
    \mathbb{E}\left\|Z - \mathbb{E}(Z)\right\|_2 &\leq K\left(\max _{i} \sqrt{\sum_{j} \mathbb{E} (Z_{ij} - p_{ij})^{2}}+\max _{j} \sqrt{\sum_{i} \mathbb{E} (Z_{ij} - p_{ij})^{2}} \right. \nonumber\\
&\left.\quad\quad\quad\quad+\sqrt[4]{\sum_{i j} \mathbb{E} (Z_{ij} - p_{ij})^{4}}\right) \nonumber\\
&\leq K\left(\sqrt{\frac{\Upsilon S}{R}}+\sqrt{\frac{\Upsilon S}{C}}+\sqrt[4]{\Upsilon S}\right),
\end{align}
where $K>0$ is an absolute constant.
\end{proof}

\subsection{First regime}\label{Sect.3.2}
In this section, we complete the proof of Theorem \ref{theo:theo1}. 

\begin{proof}[Proof of Theorem \ref{theo:theo1}]
%
The proof proceeds by approximating factors of the matrix $M$ by easier-to-handle quantities. The main technical tools involved are concentration inequalities for $N_{i\cdot},N_{\cdot j}$ and Rafa{\l} Lata{\l}a's bound on spectral norms of random matrices (see Proposition \ref{prop:RMT} and Lemma \ref{lem:RMT}).

To start with we bring in some notations : for any $i\in [R], j\in [C]$, we let $d_i=(D_1^{-1})_{ii} = \frac{1}{N_{i\cdot}+\lambda_A}$, $d_j'=(D_2^{-1})_{jj}=\frac{1}{N_{\cdot j}+\lambda_B}$.
Let
\begin{equation*}
    L=(I_R-\wone\mathbf{1}_R^{\tran})D_1^{-1}, \quad L_0=\frac{R}{S}(I_R-\frac{1}{R}\mathbf{1}_R \mathbf{1}_R^{\tran}).
\end{equation*}
We note that $M=(I_C - \wtwo\mathbf{1}_C^{\tran})M_0$ with $M_0=D_2^{-1}Z^{\tran} LZ$. In the following, we upper bound $\| M\|_2$ through four steps:
\begin{itemize}
    \item Step 1: provide an upper bound of $\|M_0\|_2$ in terms of $\|Z^{\tran}LZ\|_2$;
    \item Step 2: approximate $Z^{\tran} LZ$ by $Z^{\tran}L_0Z$, and upper bound the spectral norm of their difference;
    \item Step 3: upper bound $\|Z^{\tran}L_0Z\|_2$ by Lemma \ref{lem:RMT}, which combined with Steps 1-2 provides an upper bound of $\|M_0\|_2$ by the triangle inequality
    \begin{equation*}
        \|Z^{\tran}LZ\|_2\leq \|Z^{\tran}L_0Z\|_2+\|Z^{\tran} LZ-Z^{\tran}L_0Z\|;
    \end{equation*}
    \item Step 4: upper bound $\|M\|_2$ using the upper bound of $\|M_0\|_2$ and the inequality
    \begin{equation*}
        \|M\|_2\leq \|I_C - \wtwo\mathbf{1}_C^{\tran}\|_2\|M_0\|_2.
    \end{equation*}
\end{itemize}

\paragraph{Step 1}
Firstly we bound the quantities $N_{i \cdot}, N_{\cdot j}$ by Hoeffding's inequality. 
We take $\psi(S)=\min\{\frac{1}{20},\frac{1}{\log(S+1)}\}$, and let $\mathcal{A}_S$ be the event that for any $i\in [R],j\in [C]$, $|N_{i\cdot}-\frac{S}{R}|\leq \frac{S}{R}\psi(S)$, $|N_{\cdot j}-\frac{S}{C}|\leq \frac{S}{C}\psi(S)$.
Using Lemma \ref{lem:rowcolumn} with $\Upsilon = \Upsilon' = 1$,
under the condition on $(\rho, \kappa)$ mentioned in the theorem, $$\lim_{S\to \infty}\mathbb{P}(\mathcal{A}_S^c) = 0.$$ In the following, we assume that the event $\mathcal{A}_S$ holds. 
Note that for any $j\in [C]$,
\begin{eqnarray*}
|d_j'-\frac{C}{S}| &\leq& \frac{|\sum_{i=1}^R Z_{ij}-\frac{S}{C}|+\lambda_B}{(\sum_{i=1}^RZ_{ij})\frac{S}{C}}    \\
 & \leq   & \frac{\frac{S}{C}\psi(S)+\lambda_B}{\frac{S^2}{2 C^2}}=\delta_2\frac{C}{S},
\end{eqnarray*}
with $\delta_2=2(\psi(S)+\frac{C}{S}\lambda_B)$. 
Hence
\begin{equation}\label{eq:D2}
    \|D_2^{-1}-\frac{C}{S}I_C\|_2 = \max_{j\in [C]}|d_j'-\frac{C}{S}|\leq \delta_2\frac{C}{S}.
\end{equation}
This leads to
\begin{align*}
    &\|M_0-\frac{C}{S}Z^\tran LZ\|_2 = \|D_2^{-1}Z^{\tran} LZ -\frac{C}{S}Z^\tran LZ\|_2\\
    &\leq \|D_2^{-1}-\frac{C}{S}I_C\|_2\|Z^\tran LZ\|_2
    \leq \delta_2 \frac{C}{S}\|Z^\tran LZ\|_2.
\end{align*}
Using the triangle inequality we obtain
\begin{equation}\label{eq:mzero}
    \|M_0\|_2\leq (1+\delta_2)\frac{C}{S}\|Z^\tran LZ\|_2.
\end{equation}

\paragraph{Step 2} 
We start by bounding $\|L-L_0\|_2$. We let $D =  \wone\mathbf{1}_R^{\tran}D_1^{-1} $ with $D_{ij}=\wione d_j$. This gives
\begin{equation}\label{eq:LL0diff}
    \|L-L_0\|_2\leq  \|D_1^{-1}-\frac{R}{S}I_R\|_2+\|D-\frac{1}{S}\mathbf{1}_R \mathbf{1}_R^{\tran}\|_2.
\end{equation}
Using a similar result as equation~\eqref{eq:D2},
\begin{equation}\label{eq:D1invRbySdiff}
    \|D_1^{-1}-\frac{R}{S}I_R\|_2 = \max_{i\in [R]}|d_i-\frac{R}{S}|\leq \delta_1\frac{R}{S}. 
\end{equation}
for  $\delta_1=2(\psi(S)+\frac{R}{S}\lambda_A)$. 
For any $i,j\in [R]$, we have
\begin{eqnarray}\label{eqnn1}
|\wione d_{j}-\frac{1}{S}| 
&\leq& |\frac{\sione d_j}{\sum_{i'=1}^{R}\sipone}-\frac{\sione d_j}{R}|+|\frac{\sione d_j}{R}-\frac{d_j}{R}|+|\frac{d_j}{R}-\frac{1}{S}| \nonumber\\
&\leq& |\sione d_j||\frac{1}{\sum_{i'=1}^{R}\sipone}-\frac{1}{R}|+\frac{d_j}{R}|\sione-1|+|\frac{d_j}{R}-\frac{1}{S}| \nonumber\\
&\leq& \frac{R}{S}(1+\delta_1)\frac{1}{R}(1+\frac{2\lambda_A R}{S}-1)+\frac{d_j}{R}\frac{\frac{2\lambda_A R}{S}}{1+\frac{2\lambda_A R}{S}}+\frac{\delta_1}{S} \nonumber\\
&\leq& \frac{4(1+\delta_1)\lambda_A R}{S^2}
+\frac{\delta_1}{S}
\leq \delta_1\frac{C_0}{S},
\end{eqnarray}
where $C_0$ is a constant. In the above derivation, we have used $|\sione| \leq 1$ and $|d_j - \frac{R}{S}| \leq \delta_{1}\frac{R}{S}$ for the first and last terms while going from the inequality in the second line to the third line. We have also used the following string of inequalities :
\begin{equation}\label{eq:sioneclosetoone}
    1 \geq s_i^{(1)}=\frac{1}{1+\frac{\lambda_A}{N_{i\cdot}}}\geq \frac{1}{1+\frac{\lambda_A}{(1-\psi(S))\frac{S}{R}}}\geq \frac{1}{1+\frac{2\lambda_A R}{S}}.
\end{equation}
By (\ref{eqnn1}), we have
\begin{equation}\label{eq:wdSdiff}
    \|D-\frac{1}{S}\mathbf{1}_R \mathbf{1}_R^{\tran}\|_2\leq\|D-\frac{1}{S}\mathbf{1}_R \mathbf{1}_R^{\tran}\|_{F}=\sqrt{\sum_{i,j\in [R]}(D_{ij}-\frac{1}{S})^2}\leq \delta_1C_0 \frac{R}{S}.
\end{equation}
Plugging in the results from equations~\eqref{eq:D1invRbySdiff} and~\eqref{eq:wdSdiff} to~\eqref{eq:LL0diff}, we obtain
\begin{equation*}
    \|L-L_0\|_2\leq \delta_1(1+C_0)\frac{R}{S}.
\end{equation*}

Let $\mathcal{B}_S$ be the event that $\|Z\|,\|Z^{\tran}\|\leq \frac{2S}{\sqrt{RC}}$. By Lemma~\ref{lem:Zeigen}, 
\begin{equation*}
    \lim_{S\rightarrow \infty}\mathbb{P}(\mathcal{B}_S^c)=0.
\end{equation*}
Below we assume that $\mathcal{B}_S$ holds.
We have
\begin{eqnarray}\label{eq:ztlzztl0z}
    &&\|Z^\tran LZ-Z^\tran L_0Z\|_2\leq\|Z^\tran\|_2  \|L-L_0\|_2 \|Z\|_2\nonumber\\
    &&\leq (1+C_0)\delta_1\frac{R}{S}\frac{4S^2}{RC}  
    \leq C_0'\delta_1\frac{S}{C},
\end{eqnarray}
where $C_0'$ is a constant.

\paragraph{Step 3}
Using the fact that $\mathbb{E}(Z) = k\mathbf{1}_R\mathbf{1}_C^{\tran}$ for some constant $k$ and $L_0\bone_R = \boldsymbol{0},$ we deduce that $$Z^{\tran} L_0 Z = (Z-\mathbb{E}(Z))^{\tran} L_0(Z-\mathbb{E}(Z)).$$
We also note
\begin{align}\label{eq:L0Z}
    \Vert (Z-\mathbb{E}(Z))^{\tran}  L_0 (Z-\mathbb{E}(Z)) \Vert_2 &= \frac{S}{R}\Vert L_0(Z-\mathbb{E}(Z)) \Vert_2^2 \quad \mbox{using $L_0^2 = \frac{R}{S}L_0$} \nonumber\\
   \Vert L_0(Z-\mathbb{E}(Z)) \Vert_2 &\leq \frac{R}{S} \Vert Z-\mathbb{E}(Z)\Vert_2 \quad \mbox{because $\Vert L_0 \Vert_2 = \frac{R}{S}$}
\end{align}
By Lemma \ref{lem:RMT} (with $\Upsilon=1$), we have
\begin{equation}\label{eq:ZminusEZ}
    \mathbb{E}\Vert Z-\mathbb{E}(Z)\Vert_2\leq K\left(\sqrt{\frac{S}{R}}+\sqrt{\frac{S}{C}}+\sqrt[4]{S}\right),
\end{equation}
where $K>0$ is an absolute constant. Plugging the inequality of~\eqref{eq:L0Z} into  equation~\eqref{eq:ZminusEZ} we obtain, $$\mathbb{E}\left\|L_0\left(Z - \mathbb{E}(Z)\right)\right\|_2 \leq K\left(\sqrt{\frac{R}{S}}+\frac{R}{S^{0.5}C^{0.5}}+\frac{R}{S^{0.75}}\right)$$
We take $\epsilon=\frac{1}{\log(S+1)}$.
By Markov's inequality, 
\begin{align*}
\mathbb{P}\left( \left\|L_0\left(Z - \mathbb{E}(Z)\right)\right\|_2 > \frac{1}{\epsilon}\mathbb{E}\left\|L_0\left(Z - \mathbb{E}(Z)\right)\right\|_2 \right) &< \epsilon\\
\Longrightarrow \mathbb{P}\left( \left\|L_0\left(Z - \mathbb{E}(Z)\right)\right\|_2 > \frac{K}{\epsilon}\left(\sqrt{\frac{R}{S}}+\frac{R}{S^{0.5}C^{0.5}}+\frac{R}{S^{0.75}}\right) \right) &< \epsilon\\
\Longrightarrow \mathbb{P}\left( \frac{S}{R}\Vert L_0(Z-\mathbb{E}(Z)) \Vert_2^2 > 3(\frac{K}{\epsilon})^2\max\lbrace 1 ,\frac{R}{C},\frac{R}{S^{0.5}}\rbrace \right) &< \epsilon\\
\Longrightarrow \mathbb{P}\left( \Vert Z^{\tran}L_0 Z \Vert_2 > 3(\frac{K}{\epsilon})^2\max\lbrace 1 ,\frac{R}{C},\frac{R}{S^{0.5}}\rbrace \right) &< \epsilon
\end{align*}
Let $\mathcal{D}_\epsilon$ be the event that $\Vert Z^{\tran}L_0 Z \Vert_2 \leq 3(\frac{K}{\epsilon})^2\max\lbrace 1 ,\frac{R}{C},\frac{R}{S^{0.5}}\rbrace$.
In view of the inequality~\eqref{eq:ztlzztl0z}, we obtain on 
$\mathcal{A}_S\cap\mathcal{B}_S\cap\mathcal{D}_\epsilon$
,
\begin{equation*}
    \Vert Z^{\tran}LZ \Vert_2 \leq C_0'\delta_1\frac{S}{C} +3 (\frac{K}{\epsilon})^2\max\lbrace 1, \frac{R}{C}, \frac{R}{S^{0.5}} \rbrace. 
\end{equation*}
From equation~\eqref{eq:mzero}, on 
$\mathcal{A}_S \cap\mathcal{B}_S\cap \mathcal{D}_\epsilon$
we get 
\begin{equation}\label{eq:nearzeroMtwonorm}
\Vert M_0 \Vert_2 < (1+\delta_2)\bigl(C_0'\delta_1 + 3(\frac{K}{\epsilon})^2\max\lbrace \frac{C}{S}, \frac{R}{S}, \frac{RC}{S^{3/2}}\rbrace\bigr)
\end{equation}
The quantity on the right hand side goes to $0$ when $\rho + \kappa <\frac{3}{2}$, which is implied by the condition $\max\{\rho+\frac{1}{2}\kappa, \kappa+\frac{1}{2}\rho\}<1$. 

\paragraph{Step 4} 
For the last step, we note that
\begin{equation}\label{eq:M0Mdiff}
\|M\|_2 \leq \|I_C - \wtwo\mathbf{1}_C^{\tran}\|_2\|M_0\|_2
\end{equation}
Similar to the inequality in ~\eqref{eq:sioneclosetoone},
\begin{equation}\label{eq:sjtwoclosetoone}
    1 \geq s_j^{(2)}=\frac{1}{1+\frac{\lambda_B}{N_{\cdot j}}}\geq \frac{1}{1+\frac{\lambda_B}{(1-\psi(S))\frac{S}{C}}}\geq \frac{1}{1+\frac{2\lambda_B C}{S}} .
\end{equation}
Noting that $\wtwo = \frac{\sjtwo}{\sum_{j'=1}^{C}\sjptwo}$, we obtain for any $j\in [C]$,
\begin{equation}\label{eq:woneclose}
    \frac{1}{C} \frac{1}{1+\frac{2\lambda_B C}{S}}\leq \wjtwo \leq \frac{1}{C} (1+\frac{2\lambda_B C}{S}).
\end{equation}
Therefore, there exists $\delta(S)\rightarrow 0$, such that
\begin{equation}\label{eq:wjtwoonebyC}
    |\wjtwo-\frac{1}{C}|\leq \delta(S)\frac{1}{C}.
\end{equation}
Observe that 
\begin{align*}
    \|I_C - \wtwo\mathbf{1}_C^{\tran}\|_2 &\leq \|I_C - \frac{1}{C}\mathbf{1}_C\mathbf{1}_C^{\tran}\|_2+ \|\wtwo\mathbf{1}_C^{\tran} - \frac{1}{C}\mathbf{1}_C\mathbf{1}_C^{\tran}\|_2\\
    &\leq 1 + \|\wtwo\bone_C^{\tran} - \frac{1}{C}\mathbf{1}_C\mathbf{1}_C^{\tran}\|_F.
\end{align*} 
Using~\eqref{eq:wjtwoonebyC}, the second term can be made arbitrarily small. Plugging this into the inequality~\eqref{eq:M0Mdiff}, we obtain $\Vert M \Vert_2 \to 0$ on $\mathcal{A}_S \cap\mathcal{B}_S\cap \mathcal{D}_\epsilon$. Noting that $\lim_{S\to \infty}\mathbb{P}(\mathcal{A}_S \cap \mathcal{B}_S \cap \mathcal{D}_\epsilon) = 1$, we conclude the proof.

\end{proof}

\subsection{Second regime}\label{Sect.3.3}
In this section, we give the proof of Theorem \ref{theo:theo2}. We note that Theorem~\ref{theo:theo1} is a special case of Theorem~\ref{theo:theo2} with $\Upsilon=1$. However, the simpler case illustrated in Theorem~\ref{theo:theo1} sets up the field for the extensions in Theorems~\ref{theo:theo2} and~\ref{theo:theo3}.

\begin{proof}[Proof of Theorem \ref{theo:theo2}]

The proof of Theorem \ref{theo:theo2} involves upper bounding the spectral norm of a conjugated version $\mathcal{M}$ of the matrix $M$. We proceed by approximating $\mathcal{M}$ by a deterministic matrix $\mathcal{M}'$, which is obtained from replacing each factor of $\mathcal{M}$ by easier-to-handle quantities (see below for the precise definitions of $\mathcal{M},\mathcal{M}'$). By the triangle inequality
\begin{equation*}
    \|\mathcal{M}\|_2\leq \|\mathcal{M}-\mathcal{M}'\|_2+\|\mathcal{M}'\|_2,
\end{equation*}
we accomplish the proof in two steps. For the first step, we upper bound $\|\mathcal{M}-\mathcal{M}'\|_2$. The main technical tools involved in this step are concentration inequalities for the row and column sums $N_{i\cdot},N_{\cdot j}$ and Rafa{\l} Lata{\l}a's bound on spectral norms of random matrices (see Proposition \ref{prop:RMT} and Lemma \ref{lem:RMT}). The second step involves upper bounding $\|\mathcal{M}'\|_2$. The analysis in this step involves studying certain quadratic forms (similar to the Dirichlet form in the setting of reversible Markov chains) related to the matrix $\mathcal{M}'$.


First we set up some notations. We let $\bar{N}_{i\cdot}=\sum_{j=1}^C p_{ij}$ and $\bar{N}_{\cdot j}=\sum_{i=1}^R p_{ij}$ for any $i\in [R],j\in [C]$. We also let $\bar{D}_1=diag(\bar{N}_{i\cdot}),\bar{D}_2=diag(\bar{N}_{\cdot j})$, and $\bar{Z}=(p_{ij})$. We take $\psi(S)=\min\{\frac{1}{20},\frac{1}{\log(S+1)}\}$, and assume that $S$ is sufficiently large so that $\frac{R}{S}\lambda_A,\frac{C}{S}\lambda_B\leq \frac{1}{20}$ (note that $\rho,\kappa<1$ by the condition~\eqref{regime}).

We define 
\begin{equation*}
    \mathcal{M}=D_2^{\frac{1}{2}}M D_2^{-\frac{1}{2}},
\end{equation*}
\begin{equation*}
   \mathcal{M}'=(I_C-\frac{1}{C}\bar{D}_2^{\frac{1}{2}}\mathbf{1}_C \mathbf{1}_C^{\tran}\bar{D}_2^{-\frac{1}{2}})(\bar{D}_2^{-\frac{1}{2}}\bar{Z}^{\tran} \bar{D}_1^{-\frac{1}{2}}) (I_R-\frac{1}{R}\bar{D}_1^{\frac{1}{2}}\mathbf{1}_R\mathbf{1}_R^{\tran}\bar{D}_1^{-\frac{1}{2}})(\bar{D}_1^{-\frac{1}{2}}\bar{Z}\bar{D}_2^{-\frac{1}{2}}).
\end{equation*}
Note that $\rho(M)=\rho(\mathcal{M})\leq \Vert \mathcal{M}\Vert_2\leq \Vert \mathcal{M}-\mathcal{M}'\Vert_2+\Vert \mathcal{M}'\Vert_2$. In the following, we bound $\Vert \mathcal{M}-\mathcal{M}'\Vert_2$ and $\Vert \mathcal{M}'\Vert_2$.

\paragraph{Step 1} For this step, we provide an upper bound on $\Vert \mathcal{M}-\mathcal{M}' \Vert_2$.

We first bound the quantities $N_{i\cdot}$ and $N_{\cdot j}$ for $i\in [R],  j\in [C]$. By Hoeffding's inequality, for any $i\in [R], j\in [C]$,
\begin{equation}
    \mathbb{P}\Bigl(|N_{i\cdot}-\bar{N}_{i\cdot}|\geq \frac{S}{R}\psi(S)\Bigr)\leq 2e^{-\frac{2S^2\psi(S)^2}{CR^2}},
\end{equation}
\begin{equation}
    \mathbb{P}\Bigl(|N_{\cdot j}-\bar{N}_{\cdot j}|\geq \frac{S}{C}\psi(S)\Bigr)\leq 2e^{-\frac{2S^2\psi(S)^2}{RC^2}}.
\end{equation}
    
   We denote by ${A}_S$ be the event that for any $i\in [R], j\in[C]$, $|N_{i\cdot}-\bar{N}_{i\cdot}|\leq \frac{S}{R}\psi(S)$, $|N_{\cdot j}-\bar{N}_{j\cdot}|\leq \frac{S}{C}\psi(S)$. By the union bound we obtain
\begin{equation}
    \mathbb{P}(\mathcal{A}_S^c)\leq 2 R e^{-\frac{2S^2\psi(S)^2}{CR^2}}
    +2 C e^{-\frac{2S^2\psi(S)^2}{R C^2}}.
\end{equation}
Hence
\begin{equation}
    \lim_{S\rightarrow\infty}\mathbb{P}(\mathcal{A}_S^c)=0.
\end{equation}

In the following we assume that $\mathcal{A}_S$ holds. We let $\delta_1(S)=3(\psi(S)+\frac{R}{S}\lambda_A)$ and $\delta_2(S)=3(\psi(S)+\frac{C}{S}\lambda_B)$. Note that by our assumption, $\delta_1(S),\delta_2(S)\leq \frac{1}{2}$. For any $i\in [R]$,
\begin{equation*}
 |\frac{1}{N_{i\cdot}+\lambda_A}-\frac{1}{\bar{N}_{i\cdot}}|
\leq \frac{|N_{i\cdot}-\bar{N}_{i\cdot}|+\lambda_A}{(N_{i\cdot}+\lambda_A)\bar{N}_{i\cdot}}\leq \frac{\frac{S}{R}\psi(S)+\lambda_A}{(\frac{S}{R})^2(1-\psi(S))} \leq  \delta_1(S) \frac{R}{S}.
\end{equation*}
Similarly, for any $j\in   [C]$,
\begin{equation*}
    |\frac{1}{N_{\cdot j}+\lambda_B}-\frac{1}{\bar{N}_{\cdot j}}|\leq \delta_2(S)\frac{C}{S}.
\end{equation*}
Hence we have that
\begin{equation}
   \Vert D_1^{-1}-\bar{D_1}^{-1}\Vert_2\leq \delta_1(S)\frac{R}{S},\quad \Vert D_2^{-1}-\bar{D_2}^{-1}\Vert_2\leq \delta_2(S)\frac{C}{S}.
\end{equation}
Now note that
\begin{eqnarray}
  |\sqrt{N_{i\cdot}+\lambda_A}-\bar{N}_{i\cdot}|  \leq\frac{|N_{i\cdot}-\bar{N}_{i\cdot}|+\lambda_A}{\sqrt{N_{i\cdot}+\lambda_A}+\sqrt{\bar{N}_{i\cdot}}}
  \leq \frac{\frac{S}{R}\psi(S)+\lambda_A}{\sqrt{\frac{S}{R}}}= \frac{\delta_1(S)}{3} \sqrt{\frac{S}{R}},
\end{eqnarray}
\begin{eqnarray}
    && |\frac{1}{\sqrt{N_{i\cdot}+\lambda_A}}-\frac{1}{\sqrt{\bar{N}_{i\cdot}}}|= \frac{|\sqrt{N_{i\cdot}+\lambda_A}-\sqrt{\bar{N}_{i\cdot}}|}{\sqrt{N_{i\cdot}+\lambda_A}\sqrt{\bar{N}_{i\cdot}}} \nonumber\\
  &\leq&  \frac{\frac{\delta_1(S)}{3}\sqrt{\frac{S}{R}}}{\sqrt{(\frac{S}{R})^2(1-\psi(S))}}
  \leq \delta_1(S) \sqrt{\frac{R}{S}}.
\end{eqnarray}
Hence
\begin{equation}
    \Vert D_1^{\frac{1}{2}}-\bar{D}_1^{\frac{1}{2}}\Vert_2\leq  \delta_1(S)\sqrt{\frac{S}{R}},\quad  \Vert D_1^{-\frac{1}{2}}-\bar{D}_1^{-\frac{1}{2}}\Vert_2\leq    \delta_1(S)\sqrt{\frac{R}{S}}.
\end{equation}
Similarly,
\begin{equation}
    \Vert D_2^{\frac{1}{2}}-\bar{D}_2^{\frac{1}{2}}\Vert_2\leq  \delta_2(S)\sqrt{\frac{S}{C}},\quad  \Vert D_2^{-\frac{1}{2}}-\bar{D}_2^{-\frac{1}{2}}\Vert_2\leq    \delta_2(S)\sqrt{\frac{C}{S}}.
\end{equation}

Now note that $\|\bar{D_1}^{-1}\|_2\leq \frac{R}{S}$, $\|\bar{D}_2^{-1}\|_2\leq \frac{C}{S}$, $\|\bar{D}_1^{-\frac{1}{2}}\|_2\leq \sqrt{\frac{R}{S}}$, and $\|\bar{D}_2^{-\frac{1}{2}}\|_2\leq \sqrt{\frac{C}{S}}$. Hence by the above estimates, we have $\|D_1^{-1}\|_2\leq \frac{2R}{S}$, $\|D_2^{-1}\|_2\leq \frac{2C}{S}$, $\|D_1^{-\frac{1}{2}}\|_2\leq 2\sqrt{\frac{R}{S}}$, and $\|D_2^{-\frac{1}{2}}\|_2\leq 2\sqrt{\frac{C}{S}}$.

Using an argument similar to~\eqref{eq:sioneclosetoone}, for any $i\in [R]$, we have $\frac{1}{1+\frac{2\lambda_A R}{S}} \leq s_i^{(1)}\leq 1$ and for any $j\in [C]$,  $\frac{1}{1+\frac{2\lambda_B C}{S}} \leq s_j^{(2)}\leq 1.$
Following~\eqref{eq:wjtwoonebyC}, we conclude there exists $\delta_3(S)\rightarrow 0$, such that
\begin{equation*}
    |\wione-\frac{1}{R}|\leq \delta_3(S)\frac{1}{R}, \quad  |\wjtwo-\frac{1}{C}|\leq \delta_3(S)\frac{1}{C}.
\end{equation*}
As a result of the above two inequalities,
\begin{equation}\label{eq:woneonebyrdiff}
\begin{split}
     \|\frac{1}{R}\mathbf{1}_R\mathbf{1}_R^\tran-\wone\mathbf{1}_R^\tran\|_2\leq \|\frac{1}{R}\mathbf{1}_R\mathbf{1}_R^\tran-\wone\mathbf{1}_R^\tran\|_F \leq \delta_3(S),\\
    \|\frac{1}{C}\mathbf{1}_C\mathbf{1}_C^\tran-\wtwo\mathbf{1}_C^\tran\|_2\leq \|\frac{1}{C}\mathbf{1}_C\mathbf{1}_C^\tran-\wtwo\mathbf{1}_C^\tran\|_F \leq \delta_3(S).
    \end{split}
\end{equation}
We assume $\delta_3(S)\leq\frac{1}{4}$ (taking large $S$). Noting that $0 \leq \wione \leq \frac1R(1+\delta_3(S))$ and $0 \leq \wjtwo \leq \frac1C(1+\delta_3(S))$, we obtain 
\begin{equation}\label{eq:woneonerfrob}
\begin{split}
   \|I_R-\wone \mathbf{1}_R^\tran\|_2\leq \|I_R\|_2+\|\wone\mathbf{1}_R^\tran\|_2\leq 1+\|\wone\mathbf{1}_R^\tran\|_F \leq 4,\\
    \|I_C-\wtwo \mathbf{1}_C^\tran\|_2\leq \|I_C\|_2+\|\wtwo\mathbf{1}_C^\tran\|_2\leq 1+\|\wtwo\mathbf{1}_C^\tran\|_F \leq 4.
    \end{split}
\end{equation}
To simplify notations, in the rest of this proof we denote by
\begin{equation}\label{IR1}
    \dot{I}_R=I_R-\frac{1}{R} \mathbf{1}_R \mathbf{1}_R^{\tran}, \quad \dot{I}_C=I_C-\frac{1}{C} \mathbf{1}_C \mathbf{1}_C^{\tran};
\end{equation}
\begin{equation}\label{IR2}
    \tilde{I}_R=I_R-\wone  \mathbf{1}_R^{\tran}, \quad \tilde{I}_C=I_C-\wtwo      \mathbf{1}_C^{\tran}.
\end{equation}

By Lemma~\ref{lem:Zeigen} and the fact that $\pup<2$ (under the condition in the theorem), we have $\|\bar{Z}\|_2,\|\bar{Z}^{\tran}\|_2\leq \frac{3S}{\sqrt{RC}}$. Let $\mathcal{B}_S$ be the event that $\|Z\|_2,\|Z^{\tran}\|_2\leq \frac{3 S }{\sqrt{RC}}$. Using Lemma~\ref{lem:Zeigen} again, we have $\lim_{S\rightarrow\infty}\mathbb{P}(\mathcal{B}_S^c)=0$. In the following, we assume that the event $\mathcal{B}_S$ holds.

By Lemma~\ref{lem:RMT}, we have
\begin{equation*}
    \mathbb{E}[\Vert Z-\bar{Z}\Vert_2]\leq K_1\left(\sqrt{\frac{\Upsilon S}{R}}+\sqrt{\frac{\Upsilon S}{C}}+\sqrt[4]{\Upsilon S}\right),
\end{equation*}
where $K_1>0$ is an absolute constant. For any $\epsilon>0$, let $\mathcal{D}_\epsilon$ be the event that 
\begin{equation}\label{eq:ZminusZbar}
    \Vert Z-\bar{Z}\Vert_2\leq  \frac{K_1}{\epsilon} \left(\sqrt{\frac{\Upsilon S}{R}}+\sqrt{\frac{\Upsilon S}{C}}+\sqrt[4]{\Upsilon S}\right).
\end{equation}
By Markov's inequality, we have $\mathbb{P}(\mathcal{D}_\epsilon^c)\leq \epsilon$.

For the following we assume that $\mathcal{A}_S\cap\mathcal{B}_S\cap \mathcal{D}_\epsilon$ holds. Note that we have (recall the notations \eqref{IR1}-\eqref{IR2})
\begin{equation*}
   \mathcal{M}'=\bar{D}_2^{\frac{1}{2}}\dot{I}_C \bar{D}_2^{-1}\bar{Z}^{\tran} \dot{I}_R\bar{D}_1^{-1}\bar{Z}\bar{D}_2^{-\frac{1}{2}}.
\end{equation*}
Hence combining the above estimates (replacing each factor of $\mathcal{M}$ by the corresponding factor of $\mathcal{M}'$ one by one), we have that
\begin{eqnarray*}
 &&\|\mathcal{M}-\mathcal{M}'\|_2 \\
&\leq& \|D_2^{\frac{1}{2}}-\bar{D}_2^{\frac{1}{2}}\|_2\cdot\|\tilde{I}_C\|_2\cdot\|D_2^{-1}\|_2\cdot\|Z^{\tran}\|_2\cdot\|\tilde{I}_R\|_2\cdot\|D_1^{-1}\|_2\cdot\|Z\|_2\cdot\|D_2^{-\frac{1}{2}}\|_2 \\
&+& \|\bar{D}_2^{\frac{1}{2}}\|_2\cdot\|\tilde{I}_C - \dot{I}_C\|_2\cdot\|D_2^{-1}\|_2\cdot\|Z^{\tran}\|_2\cdot\|\tilde{I}_R\|_2\cdot\|D_1^{-1}\|_2\cdot\|Z\|_2\cdot\|D_2^{-\frac{1}{2}}\|_2\\
&+&\|\bar{D}_2^{\frac{1}{2}}\|_2\cdot\|\dot{I}_C\|_2\cdot\|D_2^{-1}-\bar{D}_2^{-1}\|_2\cdot\|Z^{\tran}\|_2\cdot\|\tilde{I}_R\|_2\cdot\|D_1^{-1}\|_2\cdot\|Z\|_2\cdot\|D_2^{-\frac{1}{2}}\|_2\\
&+& \|\bar{D}_2^{\frac{1}{2}}\|_2\cdot\|\dot{I}_C\|_2\cdot\|\bar{D}_2^{-1}\|_2\cdot\|Z^{\tran}-\bar{Z}^{\tran}\|_2\cdot\|\tilde{I}_R\|_2\cdot\|D_1^{-1}\|_2\cdot\|Z\|_2\cdot\|D_2^{-\frac{1}{2}}\|_2\\
&+& \|\bar{D}_2^{\frac{1}{2}}\|_2\cdot\|\dot{I}_C\|_2\cdot\|\bar{D}_2^{-1}\|_2\cdot\|\bar{Z}^{\tran}\|_2\cdot\|\tilde{I}_R - \dot{I}_R\|_2\cdot\|D_1^{-1}\|_2\cdot\|Z\|_2\cdot\|D_2^{-\frac{1}{2}}\|_2\\
&+& \|\bar{D}_2^{\frac{1}{2}}\|_2\cdot\|\dot{I}_C\|_2\cdot\|\bar{D}_2^{-1}\|_2\cdot\|\bar{Z}^{\tran}\|_2\cdot\|\dot{I}_R\|_2\cdot\|D_1^{-1}-\bar{D}_1^{-1}\|_2\cdot\|Z\|_2\cdot\|D_2^{-\frac{1}{2}}\|_2\\
&+&\|\bar{D}_2^{\frac{1}{2}}\|_2\cdot\|\dot{I}_C\|_2\cdot\|\bar{D}_2^{-1}\|_2\cdot\|\bar{Z}^{\tran}\|_2\cdot\|\dot{I}_R\|_2\cdot\|\bar{D}_1^{-1}\|_2\cdot\|Z-\bar{Z}\|_2\cdot\|D_2^{-\frac{1}{2}}\|_2\\
&+&\|\bar{D}_2^{\frac{1}{2}}\|_2\cdot\|\dot{I}_C\|_2\cdot\|\bar{D}_2^{-1}\|_2\cdot\|\bar{Z}^{\tran}\|_2\cdot\|\dot{I}_R\|_2\cdot\|\bar{D}_1^{-1}\|_2\cdot\|\bar{Z}\|_2\cdot\|D_2^{-\frac{1}{2}}-\bar{D}_2^{-\frac{1}{2}}\|_2\\
&\leq& K \left(\delta_1(S)+\delta_2(S)+\delta_3(S)+\epsilon^{-1}(S^{-\frac{1}{2}(1-\rho)}+S^{-\frac{1}{2}(1-\kappa)}+S^{-(\frac{3}{4}-\frac{1}{2}\rho-\frac{1}{2}\kappa)})\right),
\end{eqnarray*}
where $K>0$ is an absolute constant. Note that by the condition~\eqref{regime} we have $\phi(\rho,\kappa):=\min\{\frac{1}{2}(1-\rho),\frac{1}{2}(1-\kappa),\frac{3}{4}-\frac{1}{2}(\rho+\kappa)\}>0$. We take $\epsilon=S^{-\frac{1}{2}\phi(\rho,\kappa)}$. Then 
\begin{equation}
    \lim_{S\rightarrow\infty}\mathbb{P}((\mathcal{A}_S\cap\mathcal{B}_S \cap \mathcal{D}_\epsilon)^c)=0.
\end{equation}
We take $\delta(S)=K(\delta_1(S)+\delta_2(S)+\delta_3(S)+3 S^{-\frac{1}{2}\phi(\rho,\kappa)})$, and note that $\lim\limits_{S\rightarrow\infty}\delta(S)=0$. Moreover, when the event $\mathcal{A}_S \cap\mathcal{B}_S\cap \mathcal{D}_\epsilon$ holds, 
\begin{equation}
    \|\mathcal{M}-\mathcal{M}'\|_2\leq \delta(S).
\end{equation}

\paragraph{Step 2} For this step, we provide an upper bound on $\Vert \mathcal{M}'\Vert_2$.

We let
\begin{equation*}
    \mathcal{M}'_1=(I_R-\frac{1}{R}\bar{D}_1^{\frac{1}{2}}\mathbf{1}_R\mathbf{1}_R^{\tran}\bar{D}_1^{-\frac{1}{2}})(\bar{D}_1^{-\frac{1}{2}}\bar{Z}\bar{D}_2^{-\frac{1}{2}}),
\end{equation*}
\begin{equation*}
    \mathcal{M}'_2=(I_C-\frac{1}{C}\bar{D}_2^{\frac{1}{2}}\mathbf{1}_C\mathbf{1}_C^{\tran}\bar{D}_2^{-\frac{1}{2}})(\bar{D}_2^{-\frac{1}{2}}\bar{Z}^{\tran}\bar{D}_1^{-\frac{1}{2}}).
\end{equation*}

Note that
\begin{equation}
    \Vert\mathcal{M}'\Vert_2\leq \Vert \mathcal{M}_2'\Vert_2\Vert \mathcal{M}_1' \Vert_2.
\end{equation}

We bound $\|\mathcal{M}'_1\|_2$ in the following. Note that $(\mathcal{M}'_1)^\tran \mathcal{M}'_1 \bar{D}_2^{\frac{1}{2}}1_C=0$. In the following, we take any $u=(u_s)_{s \in [C]}$ such that $\sum_{s=1}^C u_s \bar{N}_{\cdot s}^{\frac{1}{2}}=0$.
By computation, we obtain that 
\begin{eqnarray}\label{eqnnn5}
 u^\tran (\mathcal{M}'_1)^\tran \mathcal{M}'_1 u &=& \sum_{s=1}^C\sum_{k=1}^{C}\frac{u_s}{\sqrt{\bar{N}_{\cdot s}}}\frac{u_k}{\sqrt{\bar{N}_{\cdot k}}}\frac{S_0}{R^2}(\sum_{l=1}^{R}\frac{p_{ls}}{\bar{N}_{l\cdot}})(\sum_{l=1}^{R}\frac{p_{lk}}{\bar{N}_{l\cdot}})\nonumber\\
 &+& \sum_{s=1}^C\sum_{k=1}^{C}\frac{u_s}{\sqrt{\bar{N}_{\cdot s}}}\frac{u_k}{\sqrt{\bar{N}_{\cdot k}}}(\sum_{l=1}^{R}\frac{p_{ls}p_{lk}}{\bar{N}_{l\cdot}}),
\end{eqnarray}
where $S_0:=\sum\limits_{i=1}^{R}\sum\limits_{j=1}^{C}p_{ij}\in [S,\Upsilon S]$.

Using the fact that 
$\sum_{s=1}^{C}u_s \bar{N}_{\cdot s}^{\frac{1}{2}}=0$, we have
\begin{eqnarray}\label{eqnnn1}
 &&\sum_{s=1}^C\sum_{k=1}^C \frac{u_s}{\sqrt{\bar{N}_{\cdot s}}} \frac{u_k}{\sqrt{\bar{N}_{\cdot k}}} \frac{S_0}{R^2}(\sum_{l=1}^R\frac{p_{ls}}{\bar{N}_{l\cdot }})(\sum_{l=1}^R\frac{p_{lk}}{\bar{N}_{l \cdot}})
  =\frac{S_0}{R^2}\left(\sum_{s=1}^C \frac{u_s}{\sqrt{\bar{N}_{\cdot s}}}(\sum_{l=1}^R \frac{p_{ls}}{\bar{N}_{l\cdot }})\right)^2 \nonumber \\
  &= & \frac{S_0}{R^2}\left(\sum_{s=1}^C \frac{u_s}{\sqrt{\bar{N}_{\cdot s}}}(\sum_{l=1}^R \frac{p_{ls}}{\bar{N}_{l\cdot}}-\frac{R}{S}\bar{N}_{\cdot s})\right)^2.
\end{eqnarray}
By the fact that $\frac{S}{R}\leq \bar{N}_{l\cdot}\leq \frac{\Upsilon S}{R}$, we have for every $s \in[C]$,
\begin{equation*}
 0\geq \sum_{l=1}^R\frac{p_{ls}}{\bar{N}_{l\cdot}}-\frac{R}{S}\bar{N}_{\cdot s}\geq -(1-\frac{1}{\Upsilon})\frac{R}{S}\bar{N}_{\cdot s}.
\end{equation*}
Hence
\begin{eqnarray}\label{eqnnn2}
&&\left(\sum_{s=1}^C \frac{u_s}{\sqrt{\bar{N}_{\cdot s}}}(\sum_{l=1}^R \frac{p_{ls}}{\bar{N}_{l\cdot}}-\frac{R}{S}\bar{N}_{\cdot s})\right)^2\leq (1-\frac{1}{\Upsilon})^2\frac{R^2}{S^2}(\sum_{s=1}^C |u_s|\sqrt{\bar{N}_{\cdot s}})^2\nonumber\\
&\leq& (1-\frac{1}{\Upsilon})^2\frac{R^2}{S^2}(\sum_{s=1}^C \bar{N}_{\cdot s})(\sum_{s=1}^C u_s^2)=(1-\frac{1}{\Upsilon})^2\frac{R^2S_0}{S^2}(\sum_{s=1}^C u_s^2).
\end{eqnarray}
By combining (\ref{eqnnn1}), (\ref{eqnnn2}), and the fact that $S_0\leq \Upsilon S$ , we obtain
\begin{eqnarray}\label{eqnnn3}
  &&\sum_{s=1}^C\sum_{k=1}^C \frac{u_s}{\sqrt{\bar{N}_{\cdot s}}} \frac{u_k}{\sqrt{\bar{N}_{\cdot k}}} \frac{S_0}{R^2}(\sum_{l=1}^R\frac{p_{ls}}{\bar{N}_{l\cdot }})(\sum_{l=1}^R\frac{p_{lk}}{\bar{N}_{l \cdot}})\nonumber\\
  &\leq& (1-\frac{1}{\Upsilon})^2\frac{S_0^2}{S^2}(\sum_{s=1}^C u_s^2)
  \leq (\Upsilon-1)^2 (\sum_{s=1}^C u_s^2).
\end{eqnarray}

Let $P(s,k)=\sum_{l=1}^R\frac{p_{ls}p_{lk}}{\bar{N}_{l \cdot }}$ for any $s,k\in[C]$. We note that $P(s,k)=P(k,s)$ and $\sum_{k=1}^C P(s,k)=\bar{N}_{\cdot s}$. This gives
\begin{equation*}
    \sum_{s=1}^C  u_s^2=\sum_{s=1}^C\sum_{k=1}^C\frac{u_s^2}{\bar{N}_{\cdot s}}P(s,k)=\frac{1}{2}\sum_{s=1}^C\sum_{k=1}^C\frac{u_s^2}{\bar{N}_{\cdot s}}P(s,k)+\frac{1}{2}\sum_{s=1}^C\sum_{k=1}^C\frac{u_k^2}{\bar{N}_{\cdot k}}P(s,k).
\end{equation*}
Hence
\begin{eqnarray}\label{eq:e1}
&&-\sum_{s=1}^{C}\sum_{k=1}^{C}\frac{u_s}{\sqrt{\bar{N}_{\cdot s}}}\frac{u_k}{\sqrt{\bar{N}_{\cdot k}}}P(s,k)+\sum_{s=1}^C  u_s^2  \nonumber\\
&=& \frac{1}{2}\sum_{s=1}^{C}\sum_{k=1}^{C}P(s,k)(\frac{u_s}{\sqrt{\bar{N}_{\cdot s}}}-\frac{u_k}{\sqrt{\bar{N}_{\cdot k}}})^2.
\end{eqnarray}
Now note that for any $s,k\in [C]$,
\begin{equation*}
    P(s,k)\geq R\frac{(\frac{S}{RC})^2}{\Upsilon\frac{S}{R}}=\Upsilon^{-1}\frac{S}{C^2}\geq \Upsilon^{-1}S C^{-2}\frac{\bar{N}_{\cdot s}\bar{N}_{\cdot k}}{(\frac{\Upsilon S}{C})^2}=\frac{\bar{N}_{\cdot s}\bar{N}_{\cdot k}}{\Upsilon^3S}.
\end{equation*}
Hence by~\eqref{eq:e1} we have
\begin{eqnarray}\label{eqnnn4}
&&-\sum_{s=1}^C\sum_{k=1}^C\frac{u_s}{\sqrt{\bar{N}_{\cdot s}}}\frac{u_k}{\sqrt{\bar{N}_{\cdot k}}}P(s,k)+\sum_{s=1}^C u_s^2\nonumber\\
&\geq  & \frac{1}{2\Upsilon^3 S}\sum_{s=1}^C\sum_{k=1}^C \bar{N}_{\cdot s}\bar{N}_{\cdot k}(\frac{u_s}{\sqrt{\bar{N}_{\cdot s}}}-\frac{u_k}{\sqrt{\bar{N}_{\cdot k}}})^2\nonumber\\
&\geq& \frac{1}{\Upsilon^3 }\sum_{s=1}^C u_s^2-\frac{1}{\Upsilon^3 S}(\sum_{s=1}^C \sqrt{\bar{N}_{\cdot s}}   u_s)^2= \frac{1}{\Upsilon^3}\sum_{s=1}^C u_s^2,
\end{eqnarray}
where in the last line we have used that $S_0 \geq S$ and $\sum_{s=1}^{C}\sqrt{\ndsb}u_s = 0$.
This leads to
\begin{equation}
    \sum_{s=1}^C\sum_{k=1}^C\frac{u_s}{\sqrt{\bar{N}_{\cdot s}}}\frac{u_k}{\sqrt{\bar{N}_{\cdot k}}}P(s,k)\leq (1-\frac{1}{\Upsilon^3})\sum_{s=1}^C u_s^2.
\end{equation}
By combining (\ref{eqnnn5}), (\ref{eqnnn3}), and (\ref{eqnnn4}), we conclude that
\begin{equation}
    u^{\tran}(\mathcal{M}_1')^{\tran} \mathcal{M}_1'  u\leq (1-\frac{1}{\Upsilon^3}+(\Upsilon-1)^2)(\sum_{s=1}^C  u_s^2 ).
\end{equation}
This leads to
\begin{equation}
    \Vert\mathcal{M}'_1\Vert_2=\sqrt{\|(\mathcal{M}_1')^{\tran}\mathcal{M}_1'\|_2}\leq\sqrt{ 1-\frac{1}{\Upsilon^3}+(\Upsilon-1)^2}.
\end{equation}
The following bound for $\|\mathcal{M}'_2\|$ can be similarly obtained:
\begin{equation}
    \Vert\mathcal{M}_2'\Vert_2\leq\sqrt{ 1-\frac{1}{\Upsilon^3}+(\Upsilon-1)^2}.
\end{equation}
Hence when $1\leq \Upsilon\leq 1.52$,
\begin{equation}
    \|\mathcal{M}'\|_2\leq 1-\frac{1}{\Upsilon^3}+(\Upsilon-1)^2\leq 0.9857.
\end{equation}

Recall that on the event $\mathcal{A}_S \cap\mathcal{B}_S\cap \mathcal{D}_\epsilon$, $\|\mathcal{M}-\mathcal{M}'\|_2\leq \delta(S)$, hence $\|\mathcal{M}\|_2\leq 1-\frac{1}{\Upsilon^3}+(\Upsilon-1)^2+\delta(S)$. Noting that $\lim_{S\to\infty}\mathbb{P}(\mathcal{A}_S\cap\mathcal{B}_S \cap \mathcal{D}_\epsilon) = 1$, we obtain the conclusion of the theorem.

\end{proof}

\subsection{Third regime}\label{Sect.3.4}
In this section, we give the proof of Theorem \ref{theo:theo3}. The proof of Theorem \ref{theo:theo3} follows a similar route as that of Theorem \ref{theo:theo2}.

\begin{proof} [Proof of Theorem \ref{theo:theo3}]

We reuse the notations $\bar{D}_1, \bar{D}_2,\mathcal{M}, \mathcal{M}^{'},\cdots$ from Sect.~\ref{Sect.3.3}. 
Similar to what we did for the second regime, the proof splits into two major steps. For the first step, we upper bound $\Vert \mathcal{M}-\mathcal{M}' \Vert_2$; for the second step, we upper bound $\Vert \mathcal{M}'\Vert_2$.

\paragraph{Step 1}
We start by bounding $\Vert \mathcal{M}-\mathcal{M}' \Vert_2$.

We take $S$ sufficiently large so that $\epsilon(S)<\frac{1}{4}$. As a consequence, $\nidb \geq \frac{3}{4}\frac{S}{R}$ and $\ndjb \geq \frac{3}{4}\frac{S}{C}$ for any $i\in [R], j\in [C]$. 

We take $\psi(S)=\min\{\frac{1}{20},\frac{1}{\log(S+1)}\}$, and let $\mathcal{A}_S$ be the event that
\begin{eqnarray}\label{eq:hoeffdi}
  &&  \mid \nid - \nidb \mid \leq \frac{S}{R}\psi(S) \mbox{ for any $i\in [R]$}, \nonumber \\
  &&  \mid \ndj - \ndjb \mid \leq \frac{S}{C}\psi(S) \mbox{ for any $j\in [C]$}.
\end{eqnarray}
By an argument similar to that in Sect.~\ref{Sect.3.3}, we have
\begin{equation}
    \lim_{S\rightarrow  \infty}\mathbb{P}(\mathcal{A}_S^c)=0.
\end{equation}
For the following, we would restrict our attention to $\mathcal{A}_S$.
Recalling $\nidb \geq \frac{3}{4}\frac{S}{R}$ and $\ndjb \geq \frac{3}{4}\frac{S}{C},$ we obtain $\nid \geq \frac{3}{4}\frac{S}{R} - \frac{S}{R}\psi(S)$ and $\ndj \geq \frac{3}{4}\frac{S}{C} - \frac{S}{C}\psi(S)$. We let $\delta_1(S)=3(\psi(S)+\frac{R}{S}\lambda_A)$, $\delta_2(S)=3(\psi(S)+\frac{C}{S}\lambda_B)$.
For any $i\in [R]$,
\begin{eqnarray*}
|\frac{1}{N_{i\cdot}+\lambda_A}-\frac{1}{\nidb}| \leq \frac{|N_{i\cdot}-\nidb|+\lambda_A}{(N_{i\cdot}+\lambda_A)\nidb}    
\leq \frac{\frac{S}{R}\psi(S) + \lambda_A}{\frac{S}{R}(\frac{3}{4}-\frac{1}{20})\frac{3}{4}\frac{S}{R}}
  \leq    \delta_1(S)\frac{R}{S}.
\end{eqnarray*}
Similarly, for any $j\in [C]$,
\begin{equation*}
    |\frac{1}{N_{\cdot j}+\lambda_B}-\frac{1}{\ndjb}|\leq \delta_2(S)\frac{C}{S}.
\end{equation*}
Hence we have that
\begin{equation*}
     \|D_1^{-1}-\bar{D}_1^{-1}\|_2\leq \delta_1(S)\frac{R}{S},\quad \|D_2^{-1}-\bar{D}_2^{-1}\|_2\leq \delta_2(S)\frac{C}{S},
\end{equation*}

Now note that
\begin{equation*}
    |\sqrt{N_{i\cdot}+\lambda_A}-\bar{N}_{i\cdot}|  \leq\frac{|N_{i\cdot}-\bar{N}_{i\cdot}|+\lambda_A}{\sqrt{N_{i\cdot}+\lambda_A}+\sqrt{\bar{N}_{i\cdot}}}\leq \frac{\frac{S}{R}\psi(S)+\lambda_A}{\sqrt{\frac{S}{R}(\frac{3}{4}-\frac{1}{20})}}\leq \delta_1(S)\sqrt{\frac{S}{R}},
\end{equation*}

\begin{equation*}
    |\frac{1}{\sqrt{N_{i\cdot}+\lambda_A}}-\frac{1}{\sqrt{\nidb}}|= \frac{|\sqrt{N_{i\cdot}+\lambda_A}-\sqrt{\nidb}|}{\sqrt{N_{i\cdot}+\lambda_A}\sqrt{\nidb}}\leq 2\delta_1(S)\sqrt{\frac{R}{S}}.
\end{equation*}
Hence
\begin{equation*}
    \|D_1^{\frac{1}{2}}-\bar{D}_1^{\frac{1}{2}}\|\leq  \delta_1(S)\sqrt{\frac{S}{R}},\quad  \|D_1^{-\frac{1}{2}}-\bar{D}_1^{-\frac{1}{2}}\|\leq    2\delta_1(S)\sqrt{\frac{R}{S}}.
\end{equation*}
For any $i\in [R]$, we have 
\begin{equation*}
    1 \geq s_i^{(1)}=\frac{1}{1+\frac{\lambda_A}{N_{i\cdot}}}\geq \frac{1}{1+\frac{\lambda_A}{(\frac34-\frac{1}{20})\frac{S}{R}}}\geq \frac{1}{1+\frac{2\lambda_A R}{S}}.
\end{equation*}
Using a similar argument as in equation~\eqref{eq:woneclose}, for any $i\in [R]$,
\begin{equation*}
    \frac{1}{R} \frac{1}{1+\frac{2\lambda_A R}{S}}\leq \wione \leq \frac{1}{R} (1+\frac{2\lambda_A R}{S}).
\end{equation*}
Similarly, for any $j\in [C]$,
\begin{equation*}
    \frac{1}{C} \frac{1}{1+\frac{2\lambda_B C}{S}}\leq \wjtwo \leq \frac{1}{C} (1+\frac{2\lambda_B C}{S}).
\end{equation*}
Therefore, there exists $\delta_3(S)\rightarrow 0$, such that for any $i\in [R], j\in [C]$,
\begin{equation*}
    |\wione-\frac{1}{R}|\leq \delta_3(S)\frac{1}{R}, \quad |\wjtwo-\frac{1}{C}|\leq \delta_3(S)\frac{1}{C}.
\end{equation*}
We assume $\delta_3(S)\leq\frac{1}{4}$.
From equation~\eqref{eq:hoeffdi} we obtain $\nid \leq (\frac{5}{4} + \frac{1}{20})\frac{S}{R}$  and $\ndj \leq (\frac{5}{4} + \frac{1}{20})\frac{S}{C}.$
Lemma~\ref{lem:Zeigen} along with the bounds on $\nid$ and $\ndj$ leads to $\|Z\|,\|Z^\tran\|\leq \frac{2 S }{\sqrt{RC}}$. 

By Lemma \ref{lem:RMT}, following the argument in Sect.~\ref{Sect.3.3}, there is an absolute constant $K_1>0$, such that if we let $\mathcal{D}_\epsilon$ be the event that 
\begin{equation*}
    \|Z-\bar{Z}\|_2\leq  K_1\frac{1}{\epsilon}(\sqrt{\Upsilon}\sqrt{\frac{S}{R}}+\sqrt{\Upsilon}\sqrt{\frac{S}{C}}+\Upsilon^{\frac{1}{4}}S^{\frac{1}{4}}),
\end{equation*}
then $\mathbb{P}(\mathcal{D}_{\epsilon}^c)\leq \epsilon$.

Noting that $0 \leq \wione \leq \frac1R(1+\delta_3(S))$ and $0 \leq \wjtwo \leq \frac1C(1+\delta_3(S))$, we have
\begin{eqnarray*}
\|I_R-\wone \mathbf{1}_R^\tran\|_2\leq \|I_R\|_2+\|\wone\mathbf{1}_R^\tran\|_2\leq 1+\|\wone\mathbf{1}_R^\tran\|_F \leq 4,\\
    \|I_C-\wtwo \mathbf{1}_C^\tran\|_2\leq \|I_C\|_2+\|\wtwo\mathbf{1}_C^\tran\|_2\leq 1+\|\wtwo\mathbf{1}_C^\tran\|_F \leq 4.
\end{eqnarray*}
We also have 
\begin{equation*}
    \|\frac{1}{R}\mathbf{1}_R\mathbf{1}_R^\tran-\wone\mathbf{1}_R^\tran\|_2\leq \|\frac{1}{R}\mathbf{1}_R\mathbf{1}_R^\tran-\wone\mathbf{1}_R^\tran\|_F \leq \delta_3(S),
\end{equation*}
\begin{equation*}
    \|\frac{1}{C}\mathbf{1}_C\mathbf{1}_C^\tran-\wtwo\mathbf{1}_C^\tran\|_2\leq \|\frac{1}{C}\mathbf{1}_C\mathbf{1}_C^\tran-\wtwo\mathbf{1}_C^\tran\|_F \leq \delta_3(S).
\end{equation*}


Using all the above results and the decomposition from Sect.~\ref{Sect.3.3}, we deduce that for some $\delta(S)\rightarrow 0$ and some $\epsilon\rightarrow 0$, on $\mathcal{A}_S\cap\mathcal{D}_{\epsilon}$ we have
$\|\mathcal{M}-\mathcal{M}'\|_2\leq \delta(S)$ and $\mathbb{P}((\mathcal{A}_S \cap\mathcal{D}_{\epsilon})^c)\rightarrow 0$.

\paragraph{Step 2}
In the following, we upper bound $\|\mathcal{M}'\|_2$. Letting $\mathcal{M}_1',\mathcal{M}_2'$ be defined similarly as in Sect.~\ref{Sect.3.3}, we have
\begin{equation*}
    \|\mathcal{M}'\|_2\leq \|\mathcal{M}'_1\|_2\|\mathcal{M}'_2\|_2
\end{equation*}

We bound $\|\mathcal{M}'_1\|_2$ in the following. Note that $(\mathcal{M}'_1)^\tran \mathcal{M}'_1 \bar{D}_2^{\frac{1}{2}}1_C=0$. In the following, we take $u=(u_j)$ such that $ \sum_{j=1}^C u_j \ndjb^{\frac{1}{2}}=0$. Note that the equations (\ref{eqnnn5}) and (\ref{eqnnn1}) still hold for the third regime,
with $S_0=\sum_{l=1}^{R}\sum_{k=1}^{C}p_{lk} = \sum_{l=1}^{R}\nldb \in [S(1-\epsilon(S)),S(1+\epsilon(S))]$.\\\\
Noting that $\frac{S}{R}(1-\epsilon(S)) \leq \nldb \leq \frac{S}{R}(1+\epsilon(S))$, we obtain
\begin{equation}\label{eq:plsnldb}
 |\sum_{l=1}^{R}\frac{p_{ls}}{\nldb}-\frac{R}{S}\ndsb|\leq 2\frac{R}{S}\epsilon(S) \ndsb.
\end{equation}
Hence by (\ref{eqnnn1}) and (\ref{eq:plsnldb}),
\begin{eqnarray}\label{eq:umonpartone}
  &&\sum_{s=1}^{C}\sum_{k=1}^{C}  \frac{u_s}{\sqrt{\ndsb}}\frac{u_k}{\sqrt{\ndkb}}\frac{S_0}{R^2}(\sum_{l=1}^{R}\frac{p_{ls}}{\nldb})(\sum_{l=1}^{R}\frac{p_{lk}}{\nldb}) \nonumber\\
  &=&\frac{S_0}{R^2}(\sum_{s=1}^{C} \frac{u_s}{\sqrt{\ndsb}}(\sum_{l=1}^{R} \frac{p_{ls}}{\nldb}-\frac{R}{S}\ndsb))^2 \quad \nonumber\\
  &\leq& \frac{4\epsilon(S)^2  S_0}{S^2}(\sum_{s=1}^C |u_s|\sqrt{\ndsb})^2 
  \leq \frac{4\epsilon(S)^2 S_0}{S^2}(\sum_{s=1}^{C} \ndsb)(\sum_{s=1}^{C} u_s^2)\nonumber\\
  &\leq& \frac{4\epsilon(S)^2 S_0^2}{S^2}(\sum_{s=1}^{C} u_s^2)
  \leq 16\epsilon(S)^2  (\sum_{s=1}^{C} u_s^2)
\end{eqnarray}
Let $P(s,k)=\sum_{l=1}^{R}\frac{p_{ls}p_{lk}}{\nldb}$.
Then 
\begin{equation*}
    P(s,k) = \sum_{l=1}^{R}\frac{p_{ls}p_{lk}}{\nldb}\geq R(\frac{S}{\pup RC})^2\frac{1}{\frac{5}{4}\frac{S}{R}}\geq \frac{S}{2\Upsilon^2C^2}\geq  \frac{1}{8    \Upsilon^2 S} \ndsb \ndkb.
\end{equation*}
Note that the equation (\ref{eq:e1}) still holds for the third regime. Hence
\begin{eqnarray*}
&&-\sum_{s=1}^{C}\sum_{k=1}^{C}\frac{u_s}{\sqrt{\ndsb}}\frac{u_k}{\sqrt{\ndkb}}P(s,k)+\sum_{s=1}^{C}u_s^2\\
&\geq& \frac{1}{16\Upsilon^2 S}\sum_{s=1}^{C}\sum_{k=1}^{C} \ndsb\ndkb(\frac{u_s}{\sqrt{\ndsb}}-\frac{u_k}{\sqrt{\ndkb}})^2\\
&=& \frac{S_0}{8\Upsilon^2 S}\sum_{s=1}^{C} u_s^2-\frac{1}{8\Upsilon^2 S}(\sum_{s=1}^{C} \sqrt{\ndsb}   u_s)^2 \geq \frac{1}{16\Upsilon^2}\sum_{s=1}^{C} u_s^2
\end{eqnarray*}
where in the last line we have used that $S_0 \geq \frac{S}2$ and $\sum_{s=1}^{C}\sqrt{\ndsb}u_s = 0$.
This leads to
\begin{equation}\label{eq:umonparttwo}
    \sum_{s=1}^{C}\sum_{k=1}^{C}\frac{u_s}{\sqrt{\ndsb}}\frac{u_k}{\sqrt{\ndkb}}P(s,k)\leq (1-\frac{1}{16\Upsilon^2})\sum_{s=1}^{C} u_s^2.
\end{equation}
Plugging inequalities~\eqref{eq:umonpartone} and~\eqref{eq:umonparttwo} into~\eqref{eqnnn5}, we obtain
\begin{equation*}
    \|\mathcal{M}'_1\|_2=\sqrt{\rho((\mathcal{M}'_1)^\tran \mathcal{M}'_1)}\leq \sqrt{1-\frac{1}{16\Upsilon^2}+16\epsilon(S)^2}.
\end{equation*}
A similar bound can be obtained for $\|\mathcal{M}'_2\|_2$:
\begin{equation*}
    \|\mathcal{M}'_2\|_2\leq \sqrt{1-\frac{1}{16\Upsilon^2}+16\epsilon(S)^2}.
\end{equation*}
Hence for any $1\leq \pup < \infty$ independent of $S$, there exists $\delta_0>0$, such that for $S$ sufficiently large,
\begin{equation*}
    \|\mathcal{M}'\|_2\leq 1-\delta_0.
\end{equation*}
Note that $\|\mathcal{M}-\mathcal{M}'\|_2\leq \delta(S)$ on $\mathcal{A}_S \cap \mathcal{D}_\epsilon$. Noting that $\lim_{S\to\infty}\mathbb{P}(\mathcal{A}_S \cap \mathcal{D}_\epsilon) = 1$ and $ \delta(S) \to 0$ as $S \to \infty$, we obtain the conclusion of the theorem.  

\end{proof}


\section{Simulation studies}\label{Sect.4}

\begin{figure}
  \centering
  \includegraphics[
  width=\textwidth
  ]{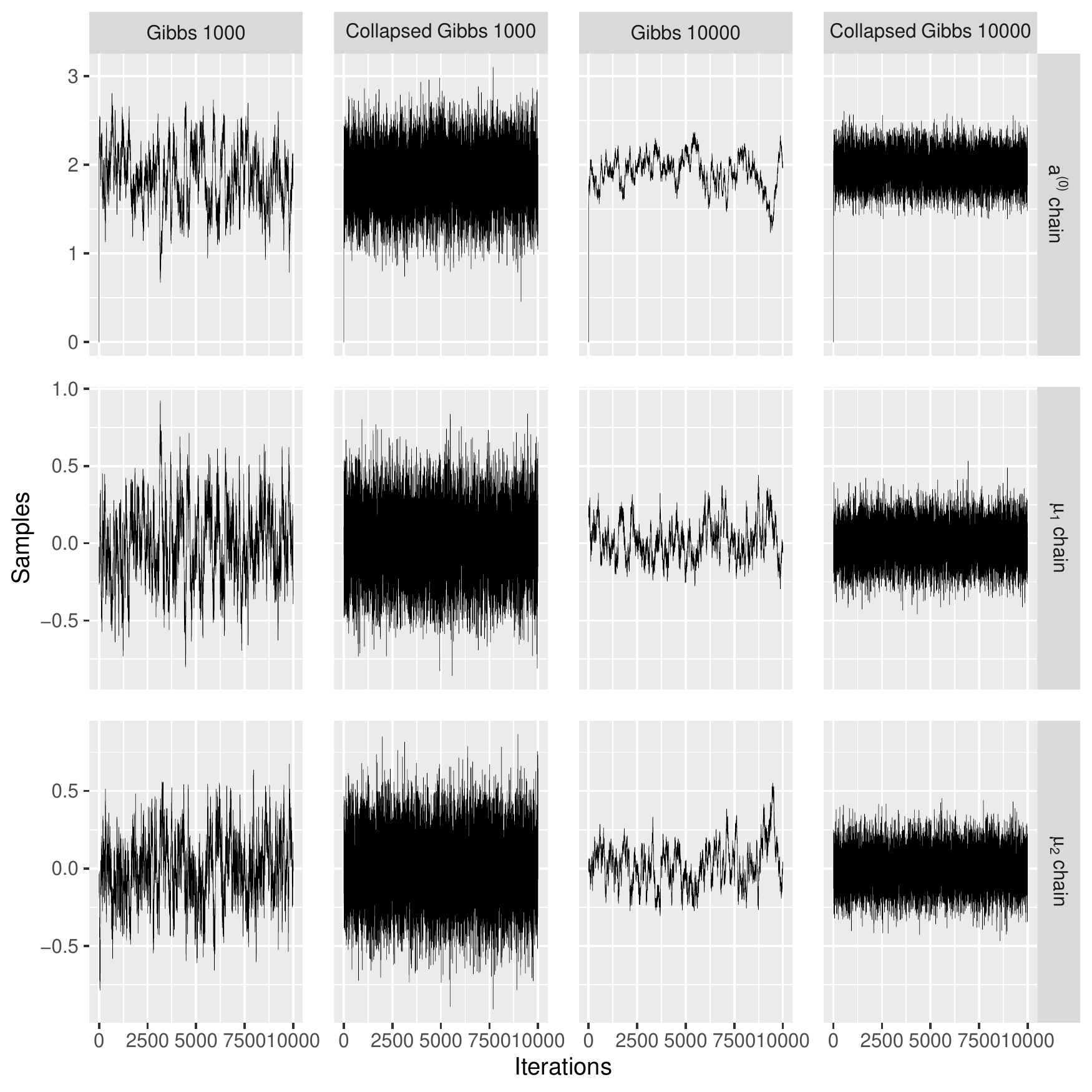}
 \caption{\label{fig:gibbscollapsedgibbstracemu1}
 Trace plots of Gibbs sampler and collapsed Gibbs sampler for mean parameters of crossed random effects model in~\eqref{eq:simmodel} for two problem sizes (S) 1000 and 10000 with $ R = \lceil S^{0.52}\rceil, C = \lceil S^{0.52}\rceil$. This indicates poor mixing of Gibbs sampler and increasing complexity with problem size.}
\end{figure}

We start off by showing some trace plots of the Gibbs sampler and the collapsed Gibbs sampler in Figure~\ref{fig:gibbscollapsedgibbstracemu1} for $10000$ iterations in simulated data for two different problem sizes.

Given {$\pij$}, we generate our observation matrix via $\zij \simind\dbern({\pij)}$.
These probabilities are first generated via
${\pij}= U_{ij}S^{1-\rho-\kappa}$ where
$U_{ij}\stackrel{\mathrm{iid}}{\sim}\dunif[1,1.52]$. Note that $1.52$ is the largest value of $\Upsilon$ up to second decimal digit for which $\phi(\Upsilon)=1-\frac{1}{\Upsilon^3}+(\Upsilon-1)^2<1$.
If $\zij = 1$, we generate $y_{ij}$ from the following model :
\begin{equation}\label{eq:simmodel}
y_{ij} = 2 + \aone_i + \atwo_{j} + e_{ij}
\end{equation}
where $\aone_i\sim \mn(0,\sigma_1^2)$ for each $1\leq i\leq R$, $a^{(2)}_j\sim \mn(0,\sigma_2^2)$ for each $1\leq j\leq C$, and $e_{ij}\sim \mn(0,\sigma_E^2)$ for every $1\leq i\leq R,1\leq j\leq C$ for $\sigma_1 = \sigma_2 = \sigma_E = 1$. 

We denote $\mu_1=\frac{1}{R}\sum_{i=1}^R a^{(1)}_i$ and $\mu_2=\frac{1}{C}\sum_{j=1}^C a^{(2)}_j$. We assume Gaussian prior on mean parameters and known precision parameters. 
Figure~\ref{fig:gibbscollapsedgibbstracemu1} indicates that the collapsed Gibbs sampler for the mean parameter in the model in~\eqref{eq:defab} with $\pup' = 1$ and $\pup = 1.52$ mixes faster than the Gibbs sampler. This is consistent with the observation made by~\cite{papaspiliopoulos2020scalable}.

\begin{figure}
  \centering
  \includegraphics[
  width=\textwidth
  ]{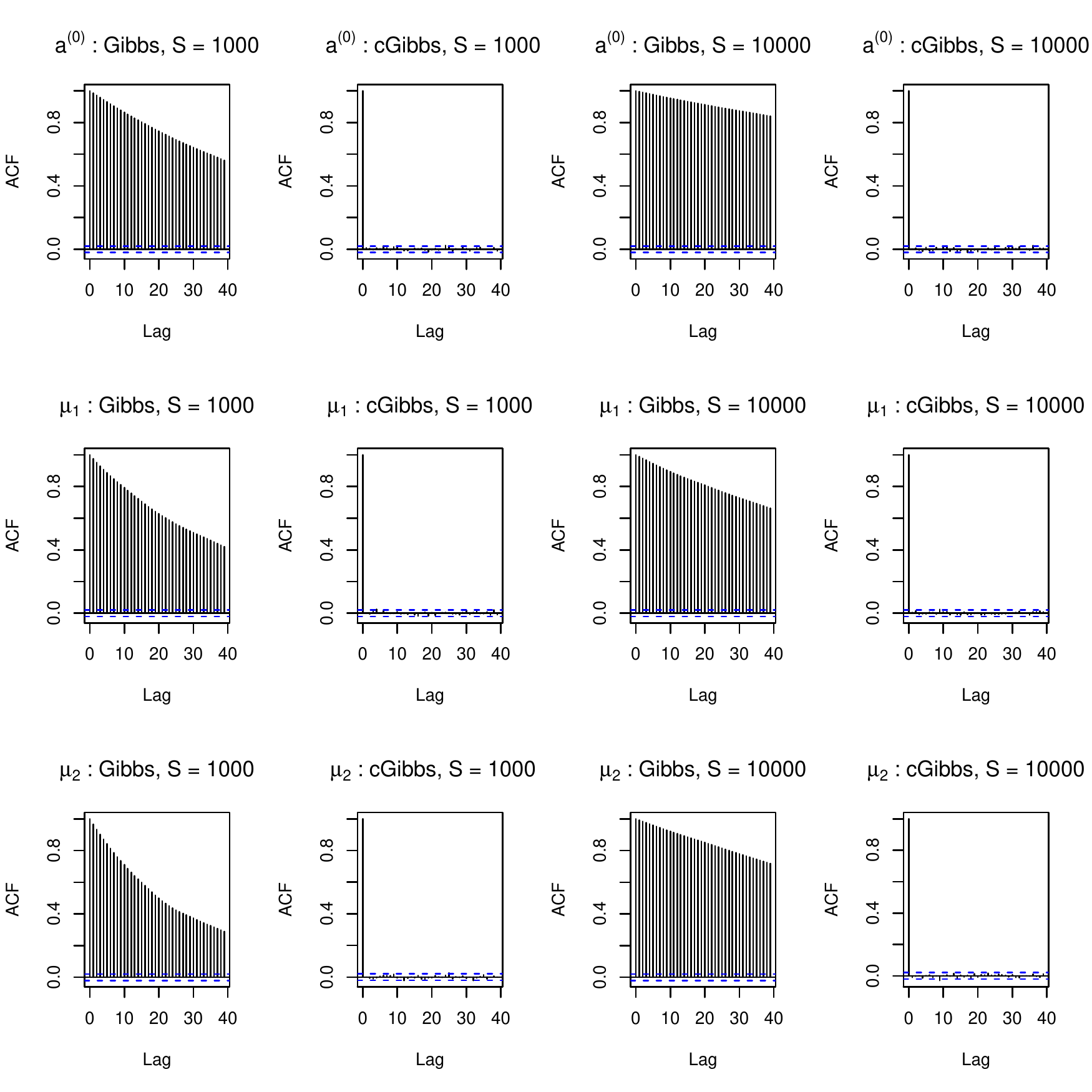}
 \caption{\label{fig:autocorrelation_plot_1000_10000}
 Autocorrelation plot of Gibbs sampler and collapsed Gibbs (cGibbs) sampler for mean parameters of crossed random effects model in~\eqref{eq:simmodel} for two problem sizes (S) 1000 and 10000 with $ R = \lceil S^{0.52}\rceil, C = \lceil S^{0.52}\rceil$. This indicates that the autocorrelation of the Gibbs sampler is significantly higher than that of the collapsed Gibbs sampler.}
\end{figure}

The trace plots of Gibbs sampler gives a sense of serial correlation of the draws. We plot the autocorrelations in Figure~\ref{fig:autocorrelation_plot_1000_10000}. The blue lines give the values beyond which the autocorrelations are statistically significantly different from zero.


We give a comparison of effective sample size for the parameters for problem size $S = 10^{\ell}$  for $\ell$ varying in $\lbrace 3, 3.25, 3.5, \cdots,5 \rbrace$. 

\begin{figure}
  \centering
  \includegraphics[width = \textwidth]{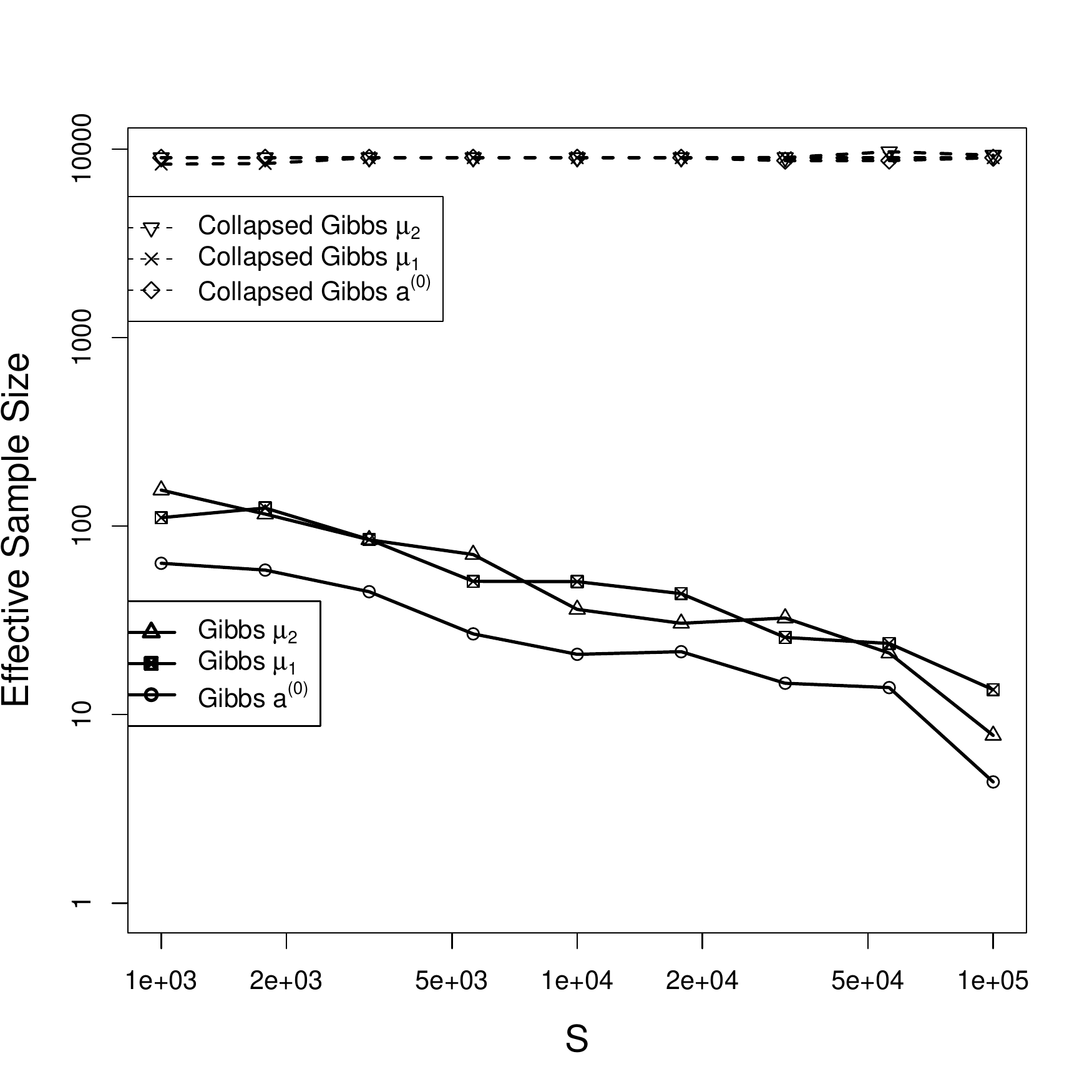}
\caption{\label{fig:effectivesamplesize0}
Effective Sample Size for mean parameters of crossed random effects of Gibbs and collapsed Gibbs sampler versus problem sizes ($S$) under the model~\eqref{eq:simmodel} with $ R = \lceil S^{0.52}\rceil, C = \lceil S^{0.52}\rceil.$
The plot has a logarithmic horizontal and vertical scale.
}
\end{figure}

Figure~\ref{fig:effectivesamplesize0} shows the effective sample size for Gibbs sampler and collapsed Gibbs sampler for $10000$ iterations after discarding first $1000$ as burn in.

Figure~\ref{fig:effectivesamplesize0} suggests that the effective sample size for Gibbs sampler goes down with problem size, while the effective sample size of its collapsed version remains constant. As noted by \cite{papaspiliopoulos2020scalable}, the mixing should depend on $(\rho,\kappa)$. We give a similar set of plots in the appendix~\ref{sec:appendix:figures} for a different value of $(\rho,\kappa)$. 

\begin{figure}
  \centering
  \centering
\includegraphics[width=0.84\textwidth]{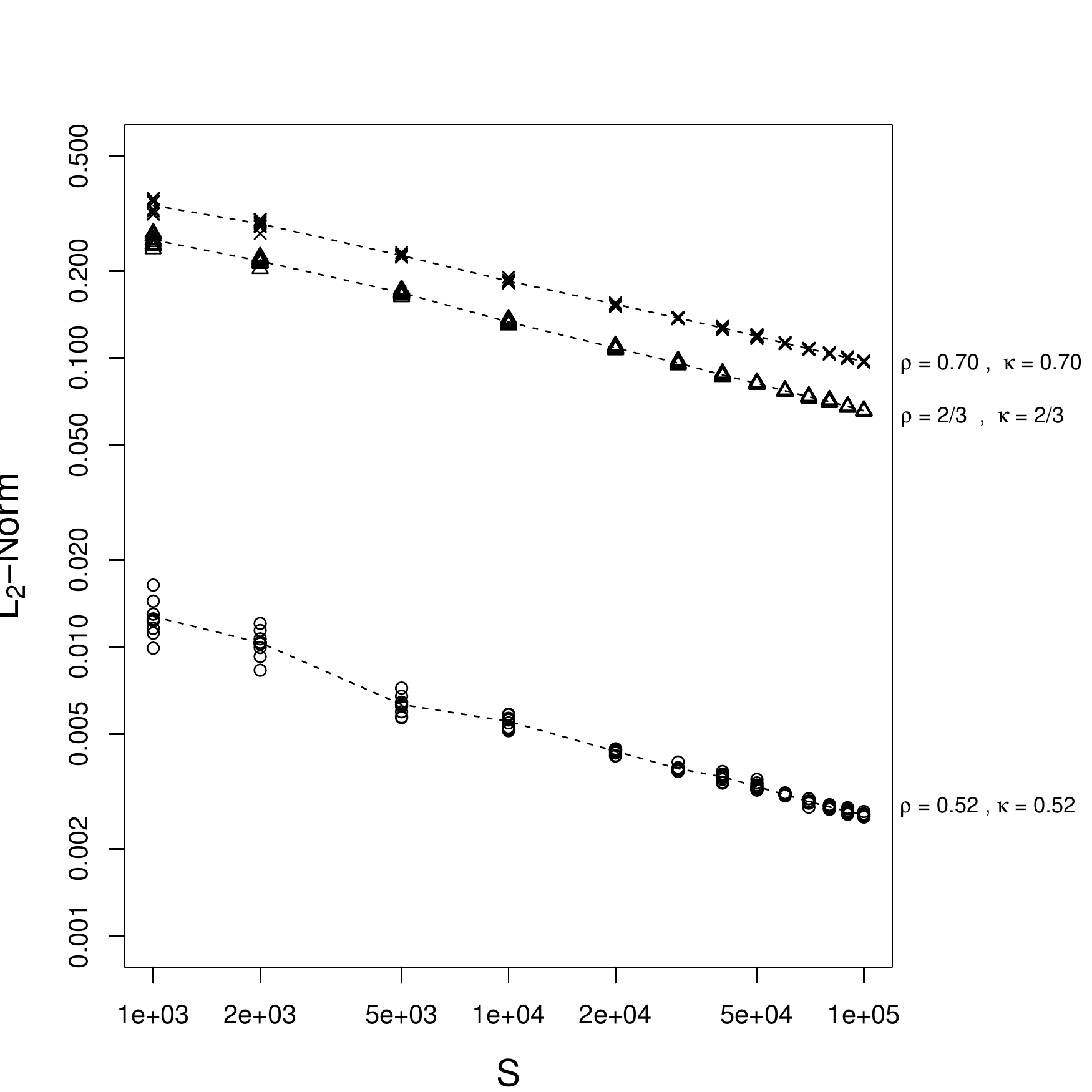}
\caption{\label{fig:5252norms}
 $L_2$ norm versus $S$
with a logarithmic vertical scale.
}
\end{figure}

In the following, we study the convergence rate of the collapsed Gibbs sampler more quantitatively through the matrix norm of the autoregression matrix (according to the results in Sect.~\ref{Sect.2}). We sample from the model multiple times at various values of $S$
and plot $\Vert M\Vert_2$ versus $S$ on a logarithmic scale.
Figure~\ref{fig:5252norms} shows the results.
We observe that $\Vert M \Vert_2$ is below $1$ and decreasing
with $S$ for all the examples $(\rho,\kappa)$ that satisfies the condition in~\eqref{regime}. We also look at $(\rho,\kappa) = (0.7,0.7)$ on the $45$ degree line that is outside
the triangle in~\eqref{regime}. The behavior of $\Vert M\Vert_2$  shows that the condition in~\eqref{regime} is a sufficient condition, but
might not be necessary.



The values of $\lambda_A$ and $\lambda_B$
appear in expressions $\nid+\lambda_A$ and $\ndj+\lambda_B$ where their
contribution is asymptotically negligible, so conservatively setting them to zero
will nonetheless be realistic for large data sets. 


\section{Application on Real Data}\label{Sect.5}

We illustrate the two samplers on some data from Stitch Fix. The same data was used in \cite{ghosh2020backfitting}.

Stitch Fix is an online service that sells clothing. First it mails the customers a sample of items, and then the customers can decide to purchase some of these items and return the others. 
Stitch Fix has provided some ratings data from their customers. Therefore, the Stitch Fix data fit naturally into the framework of crossed random effects models with missingness mechanism analyzed in this paper. Though not describing the current business (it is from 2015), this data set is useful for illustrative purposes.

Using our previous notations for crossed random effects models, we describe the details of the data set as follows. There are $N=5{,}000{,}000$ ratings in total, produced by $R=762{,}752$ customers on $C=6{,}318$ items. These correspond to $(\rho,\kappa)=(0.88,0.57)$, which does not satisfy the condition~\eqref{regime}. The response variable $y_{ij}$ is the rating of satisfaction (on a ten point scale) of customer $i$ on item $j$. Features related to the customers and the items are also included in the data set. However, as our purpose here is to study and compare large scale Gibbs sampling and collapsed Gibbs sampling, we use the following basic model, which is not necessarily the one that we would have settled on:
\begin{equation}
    y_{ij}=a^{(0)}+a^{(1)}_i+a^{(2)}_j+e_{ij}
\end{equation}
for customer $i$ and item $j$.

\begin{figure}[H]
  \centering
  \begin{subfigure}{.48\textwidth}
  \centering
  \includegraphics[scale=.32]{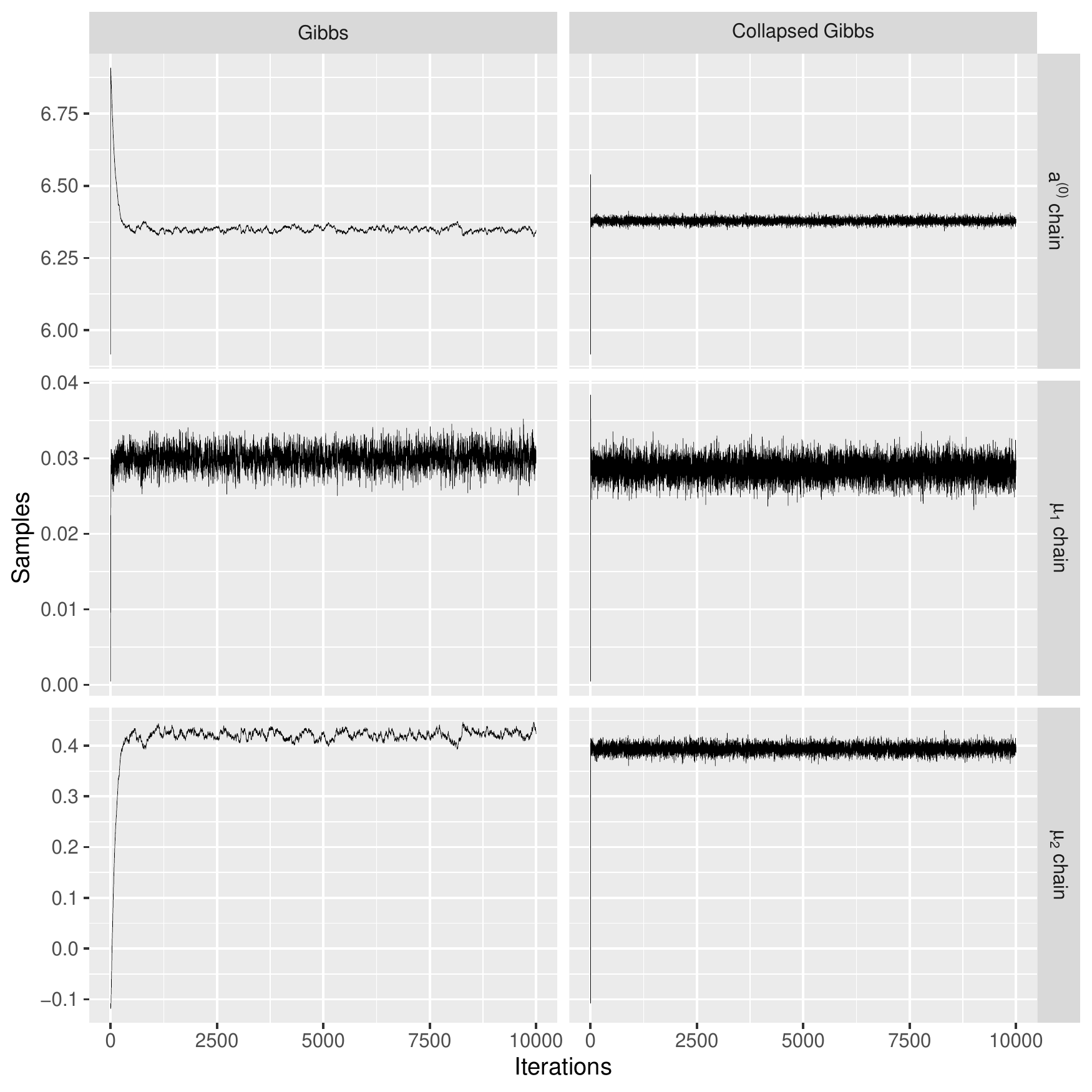}
 \end{subfigure}
\begin{subfigure}{.48\textwidth}
  \centering
\includegraphics[scale=0.32]{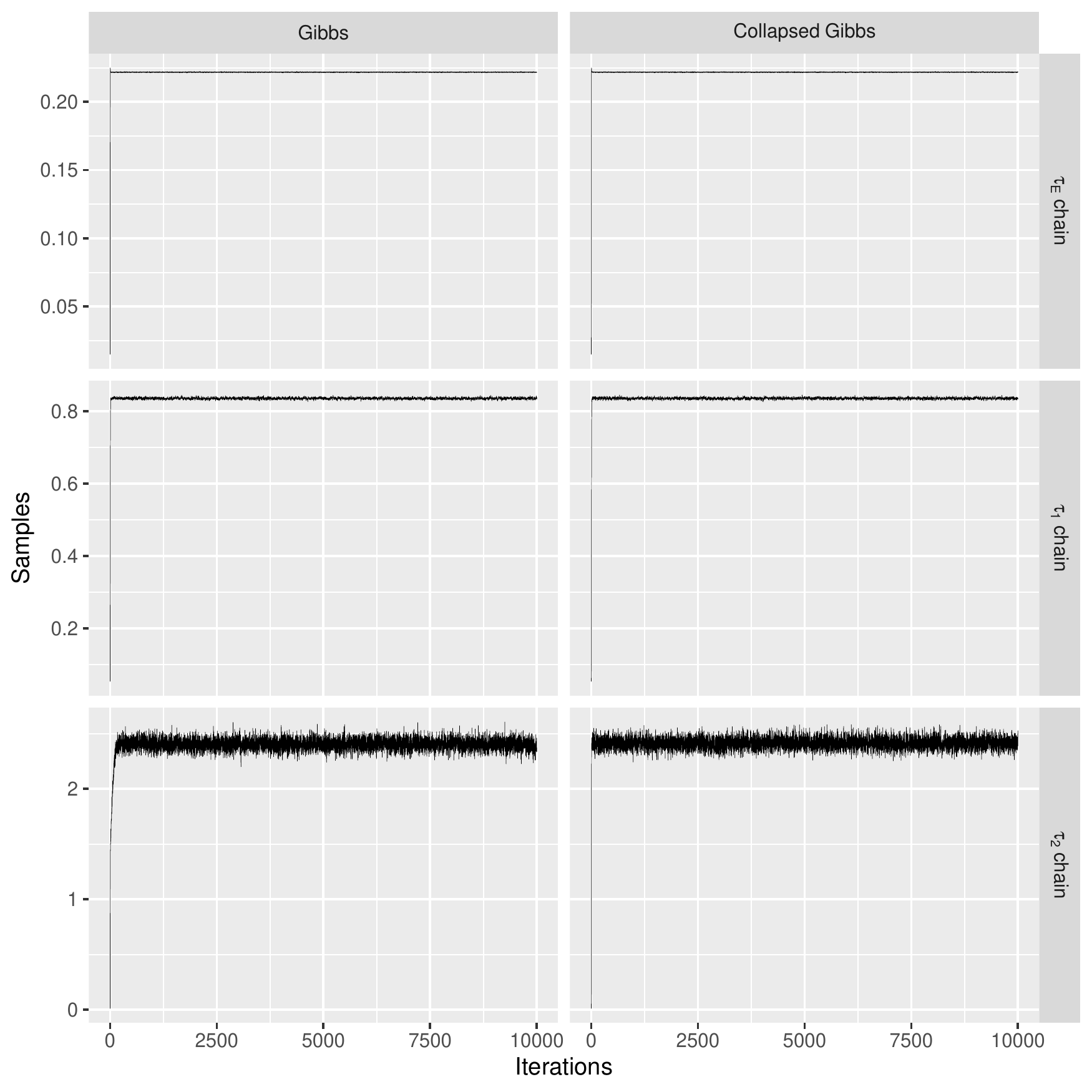}
\end{subfigure}
\caption{\label{fig:traceplotrealdata}
Trace plots of Gibbs and collapsed Gibbs sampler for crossed random effects model on Stitch Fix data. The left panel shows the trace plots for mean parameters. The right panel shows the same for precision parameters.
}
\end{figure}

We used a standard flat prior for all of the precision parameters, $\tau_E, \tau_1$ and $\tau_2$. That is, we assumed $ p(\tau^{-\frac12}) \propto 1$ for $\tau\in\{\tau_E,\tau_1,\tau_2\}$ following~\cite{papaspiliopoulos2020scalable}. We alternately sample precision parameters $\tau$ from the conditional distribution $\mathcal{L}(\tau \mid \baone,\batwo)$ and update $(\baone,\batwo)$ with the Gibbs Sampler and its collapsed version, respectively. 
In Figure~\ref{fig:traceplotrealdata}, we give the trace plots for the global mean, the means of the two random effects and the precision parameters for the Gibbs sampler and the collapsed Gibbs sampler for $10000$ iterations on Stitch Fix data. We discard the first $1000$ samples as burn-in.
We also tabulate the effective sample size in Table~\ref{table:effective_sample_size_real_data} and note that collapsed Gibbs sampler has a much larger amount of effective sample size. The similar effective sample size for $\azero$ and $\mu_2$ for Gibbs sampler might be resulting from the strong correlation of the two chains as observed in the Figure~\ref{fig:gibbsazeromu2} in the appendix.

\begin{table}[htbp]
\begin{center}
 \begin{tabular}{||c r c ||} 
 \hline
 Parameter & Gibbs & Collapsed Gibbs \\ [0.5ex] 
 \hline\hline
 $\azero$ & 67 & 9000 \\ 
 \hline
 $\mu_1$ & 1745 & 9000 \\
 \hline
 $\mu_2$ & 62 & 8566  \\
 \hline
 $\tau_E$ & 6025 & 6033 \\
 \hline
 $\tau_1$& 1746 & 1821  \\
 \hline
 $\tau_2$& 4749 & 5302\\
 \hline
\end{tabular}
\caption{Comparison of Effective Sample Size for different sampling schemes on the Stitch Fix data. Numbers are from 10000 iterations for each scheme, discarding the first 1000 samples as burn-in.
}
\label{table:effective_sample_size_real_data}
\end{center}
\end{table}

\cite{gelmshir2011} recommends an effective sample size of about $100$
posterior draws.
\cite{vatsflegjone2019} advocates even greater effective sample sizes. Clearly Gibbs sampler fails to achieve that within $10000$ iterations. 
These experiments were carried out in R on a computer
with the macOS operating system,  16 GB of memory and an Intel i7 processor.

\section{Conclusions and discussions}\label{Sect.6}
In this work, we have given sufficient conditions for finite relaxation time of a collapsed Gibbs sampler applied to crossed random effects models with unbalanced levels and a missingness mechanism. Our focus is on the Bayesian paradigm, and is therefore substantially different from the approach undertaken by \cite{ghosh2020backfitting}, in which they obtained the GLS estimate in a scalable fashion under less stringent conditions compared to the ``balanced levels'' condition. \cite{ghosh2020backfitting} studied the $L_1$ norm of their backfitting matrix. They found in practice $L_2$ norm behaves desirably for scalability compared to the $L_1$ norm. They also found the $L_2$ norm and spectral radius are in a close range. 
Studying the $L_2$ norm using the tools from random matrix theory lets us weaken the condition on the missingness mechanism that has been considered in \cite{ghosh2020backfitting}. Moreover, our proof strategy allows us to incorporate an arbitrary inhomogeneity level (measured by $\Upsilon$) for missingness pattern when the row sums and column sums are ``almost balanced'' in expectation (compared to the constraint $\Upsilon\leq 1.27$ in \cite{ghosh2020backfitting}). Such kind of results are new to the best of our knowledge. The convergence rate of the collapsed Gibbs sampler when the data satisfy the ``balanced levels'' condition was obtained by Papaspiliopoulos et al. \cite[Theorem 4]{papaspiliopoulos2020scalable}. This requires that $N_{i\cdot}=\frac{N}{R}$ for all $i=1,\cdots,R$ and $N_{\cdot j}=\frac{N}{C}$ for all $j=1,\cdots,C$. In their result, the convergence rate (denoted by $\rho_{PRZ}$) of the collapsed Gibbs sampler is expressed through the convergence rate $\rho_{aux}$ of an auxiliary process:
\begin{equation}
    \rho_{PRZ}=\frac{N\sigma_1^2}{N\sigma_1^2+R\sse}\times\frac{N\sigma_2^2}{N\sigma_2^2+C\sse}\times\rho_{\mathrm{aux}}
\end{equation}
in our notations. The auxiliary process is defined as the Gibbs sampler with stationary distribution given by $\frac{Z_{ij}}{N}$ for any $i=1,\cdots,R$ and $j=1,\cdots,C$. Each step of this auxiliary process can be described as follows: given $j\in \{1,\cdots,C\}$, sample $i\in \{1,\cdots,R\}$ with probability $\frac{Z_{ij}}{N_{\cdot j}}$; given $i$, sample $j'\in\{1,\cdots,C\}$ with probability $\frac{Z_{ij'}}{N_{i \cdot}}$, and move to $j'$. Compared to their results, our results work for several regimes with unbalanced levels, and give unconditional (i.e., without assuming finite relaxation time of any auxiliary process) convergence results. We also believe that the asymptote like the one that we study where $R,C,N$ simultaneously grow is a better description for sparse data coming from electronic commerce.

There can be multiple ways to go forward. One immediate direction this work can move forward is to obtain a characterisation of the rate of the collapsed Gibbs sampler when the number of crossed random effects is bigger than 2. Another interesting problem is when the response variable is binary and we do not have Gaussian error. We expect that the progress made here
will be useful for those problems. A further direction is to incorporate informative missingness into the model. There is usually a selection bias regarding which data points are observed, and it is expected that information from outside the current data set will be needed to account for such selection bias. We expect that our current solution will help future efforts for incorporating informative missingness (possibly by reweighting the observations).

\section*{Acknowledgements}
The authors are grateful to Brad Klingenberg and Stitch Fix for sharing some test data with us. S.G. was supported by the U.S.\ National Science Foundation under grant IIS-1837931. 
We would also like to thank Persi Diaconis, Art Owen, Omiros Papaspiliopoulos and Giacomo Zanella for several helpful discussions throughout the preparation of this manuscript.

\appendix
\section{Some proofs}\label{sec:appendix}

\subsection{Proof of Proposition~\ref{prop:collapsedgibbsmatrix}}\label{sec:proof:prop:collapsedgibbsmatrix}

Observe that
\begin{eqnarray}
&&E\left(a_{i}^{(1)}(t+1) \mid a^{(0)}(t+1), \batwo(t)\right)  \nonumber\\
&=& \sione\left(\tilde{y}_{i}^{(1)}-a^{(0)}(t+1)-\frac{1}{\nid}\sum_{j}a_{j}^{(2)}(t) Z[i, j]\right).
\end{eqnarray}
Marginalizing over $a^{(0)}(t + 1)$, we obtain
\begin{eqnarray}\label{eq:a1}
&&E\left(a_{i}^{(1)}(t+1) \mid \batwo(t)\right)\nonumber\\
&=& \sione\left(\tilde{y}_{i}^{(1)}- \mathbb{E}(a^{(0)}(t+1) \mid \batwo(t)) -\frac{1}{\nid}\sum_{j}a_{j}^{(2)}(t) Z[i, j]\right).
\end{eqnarray}
We also note that 
\begin{eqnarray}\label{eq:a2}
&&E\left(a^{(0)}(t+1) \mid \batwo(t)\right) \nonumber\\ &=&\frac{1}{\sum_{i'=1}^{R}\sipone} \sum_{i=1}^{R}\sione\left(\tilde{y}_{i}^{(1)} - \frac{1}{\nid}\sum_{j}a_{j}^{(2)}(t) Z[i, j]\right).
\end{eqnarray}
Substituting~\eqref{eq:a2} in~\eqref{eq:a1}, we conclude
\begin{align}\label{a1giva2}
E\left(a_{i}^{(1)}(t+1) \mid \batwo(t)\right) &= \frac{\sione}{\sum_{l}\slone}\sum_{i'=1}^{R}\frac{\sipone}{\nipd}\sum_{j}Z[i', j]a_{j}^{(2)}(t) \nonumber \\ 
&- \sum_{j=1}^{C}\frac{\sione Z[i,j]}{\nid}a^{(2)}_{j}(t) + C,
\end{align}
where $C$ is a constant. From the definition of $B_1$, we know that $B_1[i,j]$ is the coefficient of $a^{(2)}_j(t)$ in the conditional expectation of $a_i^{(1)}(t + 1)$ given $\batwo(t)$. Thus~\eqref{a1giva2} implies that
$$B_{1}[i,j] = \sione \left[\frac{\sum_{i'=1}^{R}Z[i',j](\sipone/\nipd)}{\sum_{l=1}^{R}\slone}-\frac{Z[i,j]}{\nid}\right].$$
Noting that $\frac{\sione}{\nid} = \frac{1}{\nid + \lambda_A}$ and $\wione = \frac{\sione}{\sum_{i'}\sipone}$, we simplify the expression as 
\begin{equation}\label{eq:b1}
B_{1}[i,j] = -\frac{Z[i,j]}{\nid+\lambda_A} + \wione\sum_{i'=1}^{R} \frac{Z[i',j]}{\nipd + \lambda_A} = -\frac{Z[i,j]}{\nid+\lambda_A} + \wione u_j
\end{equation}
for $u_j = \sum_{i'=1}^{R} \frac{Z[i',j]}{\nipd + \lambda_A}.$ Similarly, 
\begin{equation}\label{eq:b2}
B_{2}[j,i] = -\frac{Z^{\tran}[j,i]}{\ndj+\lambda_B} + \wjtwo\sum_{j'=1}^{C} \frac{Z^{\tran}[j',i]}{\ndjp + \lambda_B} = -\frac{Z^{\tran}[j,i]}{\ndj+\lambda_B} + \wjtwo l_i
\end{equation}
for $l_i = \sum_{j'=1}^{C} \frac{Z^{\tran}[j',i]}{\ndjp + \lambda_B}.$

A simple computation gives that $\batwo(t)$ is a Gaussian autoregressive process with autoregression matrix $B_2B_1.$
Combining expressions in~\eqref{eq:b1} and~\eqref{eq:b2} we obtain
\begin{eqnarray}\label{eq:b2b1}
(B_2B_1)_{js} &=& \sum_{i=1}^{R} \frac{Z^{\tran}[j,i]}{\ndj+\lambda_B} \left(\frac{Z[i,s]}{\nid+\lambda_A} - \wione u_s \right) \nonumber\\
 &-  & \wjtwo\times \sum_{i=1}^{R} l_i \left(\frac{Z[i,s]}{\nid+\lambda_A} - \wione u_s \right).
\end{eqnarray}

To analyze the convergence rate of the two component Gibbs sampler it suffices to study the convergence of one chain (See \cite{robertsahoo2001}).  Then the autoregression matrix for collapsed Gibbs sampler is obtained upon translating \eqref{eq:b2b1} in matrix notation : $$M = (I_C - \wtwo \mathbf{1}_C^{\tran})M_0.$$ By \cite[Theorem 1]{roberts1997updating}, we have
\begin{equation}
    t_{rel}=\frac{1}{1-\rho(M)}.
\end{equation}

\subsection{Proof of Lemma~\ref{lem:rowcolumn}}\label{sec:proof:lem:rowcolumn}

Letting $X \stocleq Y$ mean that $X$ is stochastically smaller than $Y$, we know that
\begin{align*}
\dbin(R, \frac{1}{\pup'}S^{1-\rho-\kappa}) &\stocleq \sum_i \zij \stocleq \dbin( R, \pup S^{1-\rho-\kappa}),\quad\text{and}\\
\dbin(C,\frac{1}{\pup'}S^{1-\rho-\kappa}) &\stocleq \sum_j \zij \stocleq \dbin( C, \pup S^{1-\rho-\kappa}).
\end{align*}
By Lemma  \ref{lem:hoeff}, if $\psi\ge0$, then
\begin{eqnarray*}
\mathbb{P}( \sum_j \zij \ge S^{1-\rho}(\pup+\psi))
&\le& \mathbb{P}\bigl( \dbin(C,\pup S^{1-\rho-\kappa}) \ge S^{1-\rho}(\pup+\psi)\bigr)\\
&\le& \exp(-2(S^{1-\rho}\psi)^2/C),
\end{eqnarray*}
\begin{eqnarray*}
\mathbb{P}\Bigl( \max_{1\le i\le R} \sum_j \zij \ge S^{1-\rho}(\pup+\psi) \Bigr) &\le R\exp(-2S^{2-\kappa-2\rho}\psi^2).
\end{eqnarray*}
Similarly,
\begin{equation*}
    \mathbb{P}\Bigl( \min_{1\le i\le R} \sum_j \zij \le S^{1-\rho}(\frac{1}{\pup'}-\psi) \Bigr) \le R\exp(-2S^{2-\kappa-2\rho}\psi^2).
\end{equation*}
Therefore for any $\psi>0$,
\begin{align*}
&\mathbb{P}\Bigl( S^{1-\rho}(\frac1{\pup'} - \psi) \le \min_{1\le i\le R} \sum_j   \zij \le \max_{1\le i\le R} \sum_j   \zij \le S^{1-\rho}(\pup + \psi) \Bigr) \\
&\geq 1-2R\exp(-2S^{2-\kappa-2\rho}\psi^2).
\end{align*}
If $\rho+\frac{1}{2}\kappa<1$, then for any fixed $\psi>0$,
\begin{align*}
\lim_{S\to \infty}\mathbb{P}\Bigl( S^{1-\rho}(\frac1{\pup'} - \psi) \le \min_{1\le i\le R} \sum_j   \zij \le \max_{1\le i\le R} \sum_j   \zij \le S^{1-\rho}(\pup + \psi) \Bigr) &= 1.
\end{align*}
Similarly, for any $\psi>0$,
\begin{align*}
&\mathbb{P}\Bigl(S^{1-\kappa} (\frac{1}{\pup'}-\psi)\le \min_{1\leq j\leq C} \sum_i \zij \le \max_{1\leq j\leq C} \sum_i \zij \le S^{1-\kappa}(\pup+\psi) \Bigr)\nonumber\\
&\geq 1-2C \exp(-2S^{2-\rho-2\kappa}\psi^2).
\end{align*}
If $\frac{1}{2}\rho+\kappa<1$, then for any fixed $\psi>0$,
\begin{align*}
\lim_{S\to\infty}\mathbb{P}\Bigl( S^{1-\kappa}(\frac{1}{\pup'}-\psi)\le \min_{1\leq j\leq C} \sum_i \zij \le \max_{1\leq j\leq C} \sum_i \zij \le S^{1-\kappa} (\pup+\psi) \Bigr)=1.
\end{align*}

\subsection{Proof of Lemma~\ref{lem:Zeigen}}\label{sec:proof:lem:eigenv}


Using the inequality $\|.\|_2 \leq \sqrt{\|.\|_1 \|.\|_\infty}$, we obtain $\|Z\|_2 \leq \sqrt{\max_i \nid \max_j \ndj}$. Using lemma~\ref{lem:rowcolumn}, under the model~\eqref{eq:defab} and the condition in~(\ref{regime}), we obtain $\|Z\|_2 \leq (\Upsilon+1) \frac{S}{\sqrt{RC}}$ with high probability.

\subsection{Some additional figures}\label{sec:appendix:figures}
Below we give the trace plot, effective sample size and autocorrelation plot for $(\rho,\kappa) = (0.36,0.66)$. Between the two factors, the mixing and effective sample size seems to poorer for the factor with lower number of levels.

\begin{figure}[H]
  \centering
  \includegraphics[
  width=\textwidth
  ]{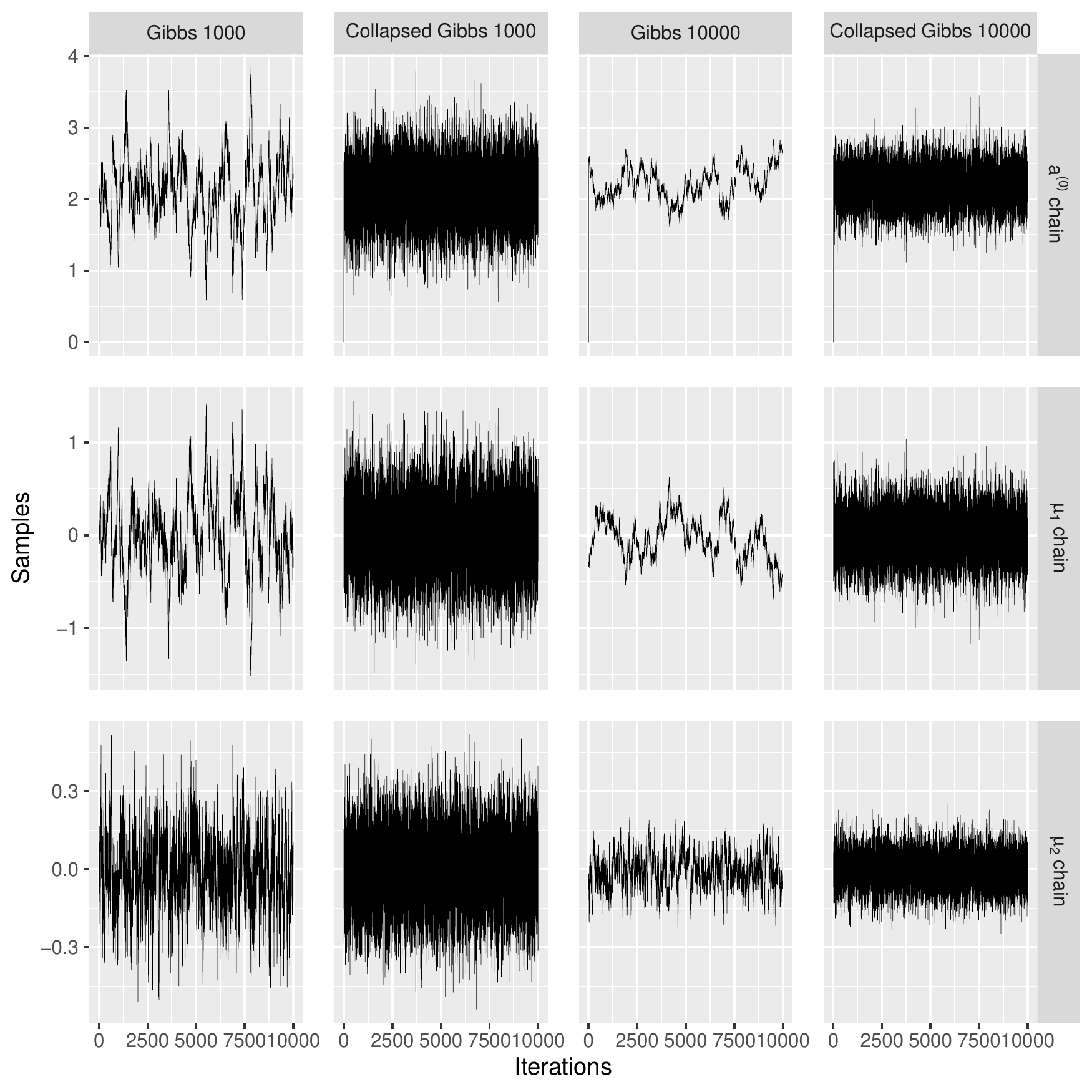}
 \caption{\label{fig:gibbscollapsedgibbstracemu13666}
 Trace plots of Gibbs sampler and collapsed Gibbs sampler for mean parameters of crossed random effects model in~\eqref{eq:simmodel} for two problem sizes (S) 1000 and 10000 with $ R = \lceil S^{0.36}\rceil, C = \lceil S^{0.66}\rceil$. This indicates poor mixing of Gibbs sampler and increasing complexity with sample size. The mixing seems to be poorer for the factor with lower number of levels.}
\end{figure}


\begin{figure}[H]
  \centering
  \includegraphics[
  width=\textwidth
  ]{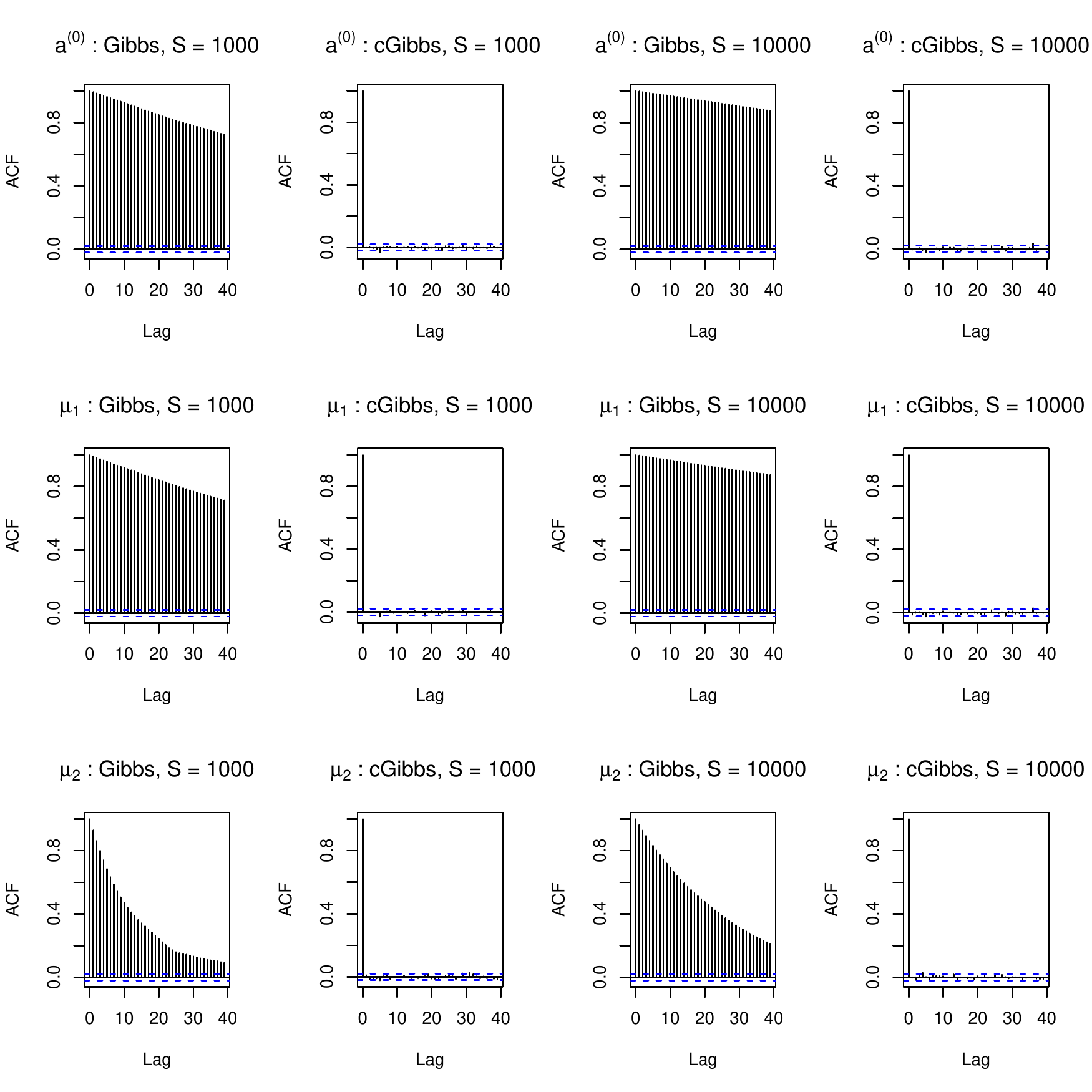}
 \caption{\label{fig:autocorrelation_plot_1000_10000_36_66}
 Autocorrelation plot of Gibbs sampler and collapsed Gibbs (cGibbs) sampler for mean parameters of crossed random effects model in~\eqref{eq:simmodel} for two problem sizes (S) 1000 and 10000 with $ R = \lceil S^{0.36}\rceil, C = \lceil S^{0.66}\rceil$. This indicates that the autocorrelation of the Gibbs sampler is significantly higher than that of the collapsed Gibbs sampler.}
\end{figure}

\begin{figure}[H]
  \centering
  \includegraphics[width = \textwidth]{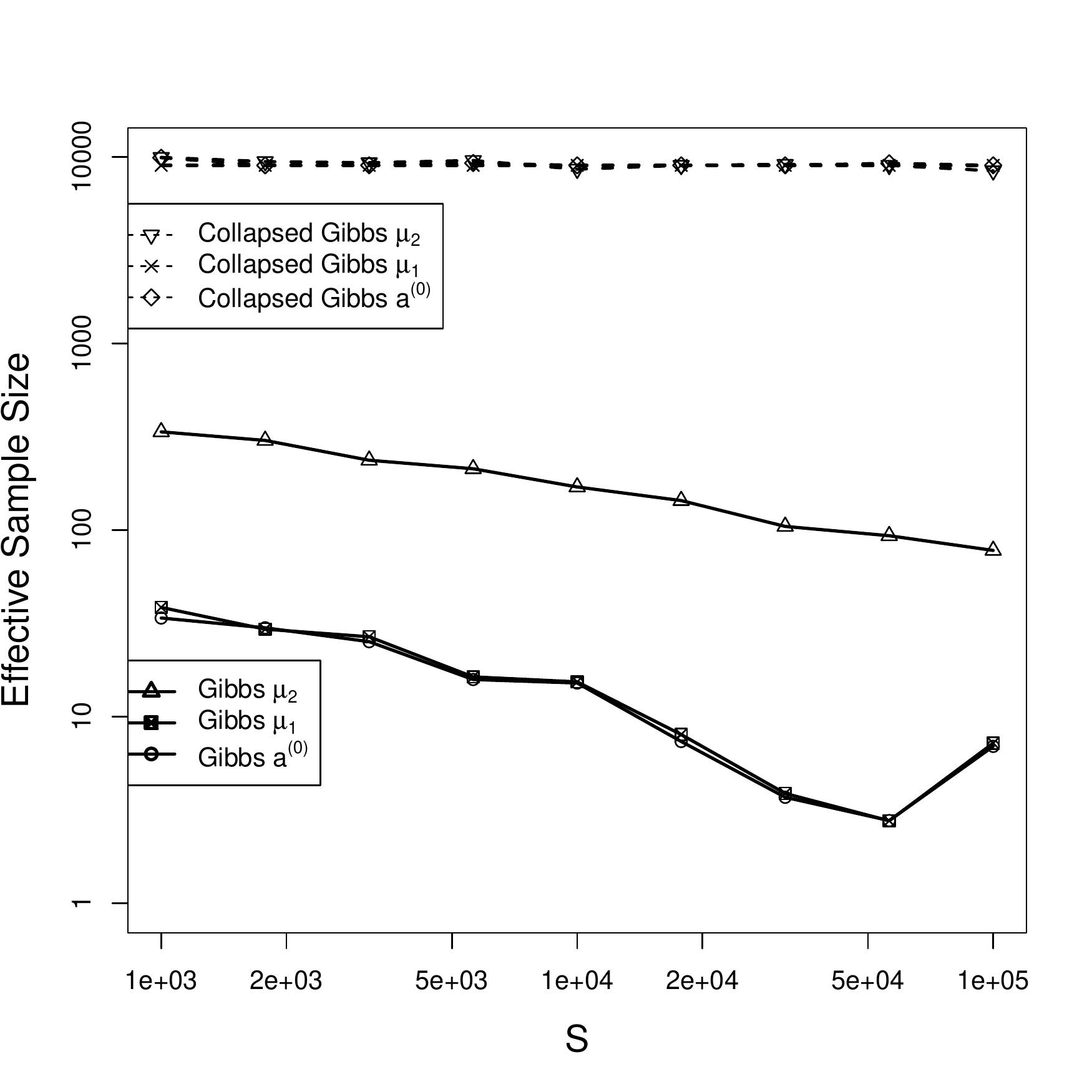}
\caption{\label{fig:effectivesamplesize}
Effective Sample Size for mean parameters of crossed random effects of Gibbs and collapsed Gibbs sampler versus problem sizes ($S$) under the model~\eqref{eq:simmodel} with $ R = \lceil S^{0.36}\rceil, C = \lceil S^{0.66}\rceil.$
The plot has a logarithmic horizontal and vertical scale.
}
\end{figure}
For the figures above and in Sect.~\ref{Sect.4} we assumed precision parameters to be known. We observe a similar behavior of trace plot, autocorrelation and effective sample size as we did for the unknown precision parameter case. For brevity, we give the trace plot for mean and precision parameters for the simulation set up in model~\eqref{eq:simmodel} with flat prior on $\tau^{-\frac12}_1$, $\tau^{-\frac12}_2$ and $\tau^{-\frac12}_E$.

\begin{figure}[H]
  \centering
  \includegraphics[
  width=\textwidth
  ]{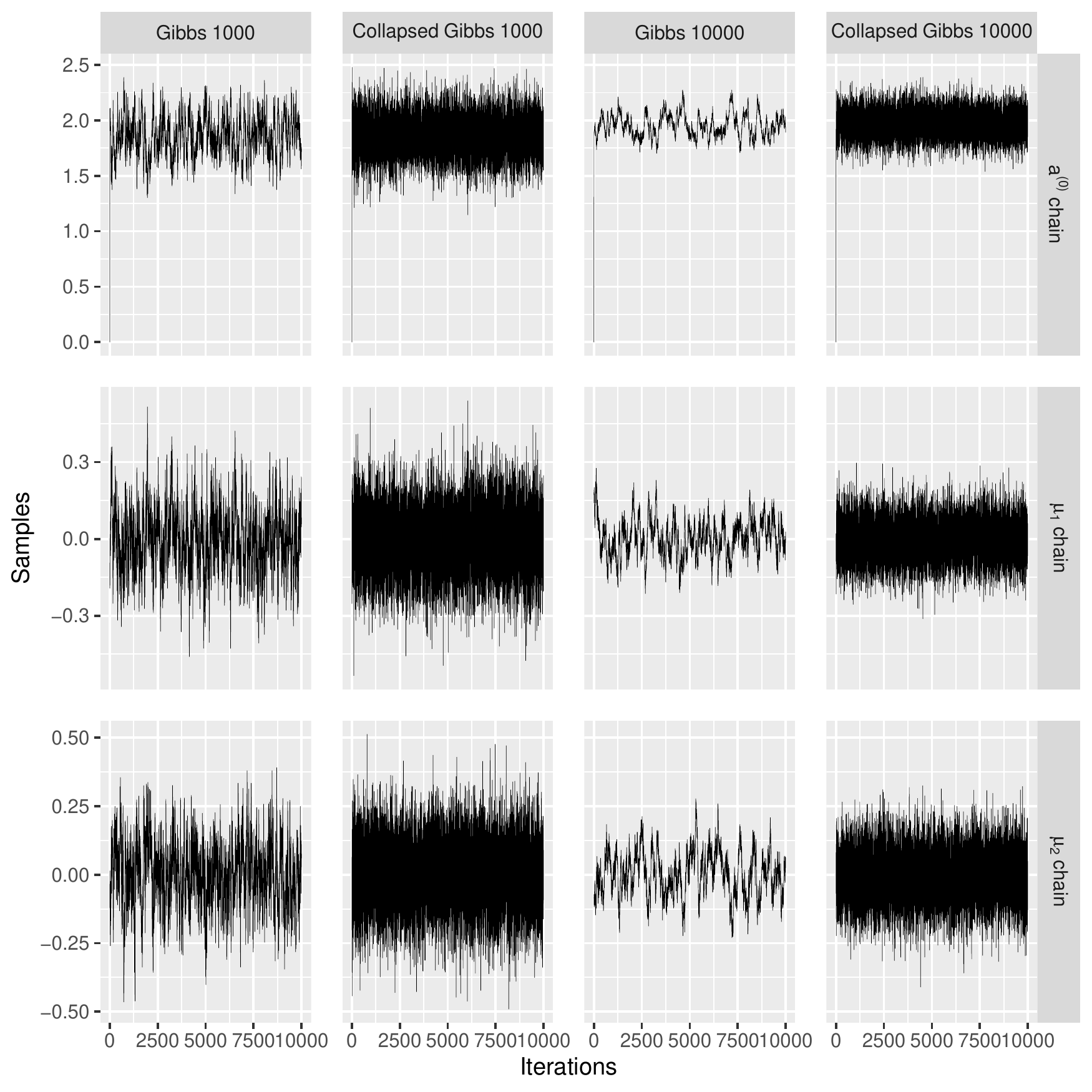}
 \caption{\label{fig:gibbscollapsedgibbstracemuunknowntau5252}
Trace plots of Gibbs sampler and collapsed Gibbs sampler for mean parameters of crossed random effects model in~\eqref{eq:simmodel} for two problem sizes (S) 1000 and 10000 with $ R = \lceil S^{0.52}\rceil, C = \lceil S^{0.52}\rceil$. This indicates poor mixing of Gibbs sampler and increasing complexity with sample size. The mixing seems to be poorer for the factor with lower number of levels.}
\end{figure}

\begin{figure}[H]
  \centering
  \includegraphics[
  width=\textwidth
  ]{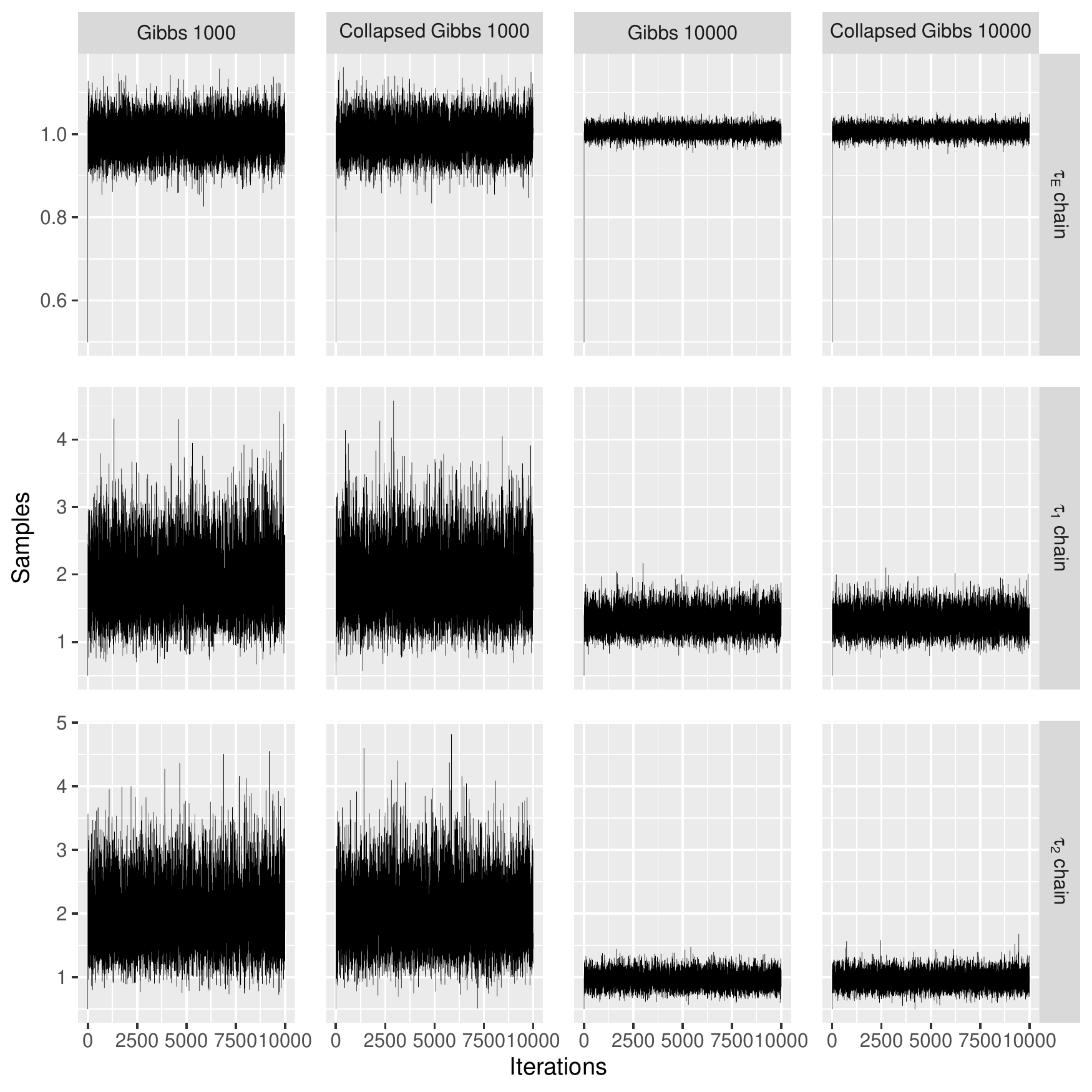}
 \caption{\label{fig:gibbscollapsedgibbstracetauunknowntau5252}
 Trace plots of Gibbs sampler and collapsed Gibbs sampler for precision parameters of crossed random effects model in~\eqref{eq:simmodel} for two problem sizes (S) 1000 and 10000 with $ R = \lceil S^{0.52}\rceil, C = \lceil S^{0.52}\rceil$. This indicates similar mixing of Gibbs sampler and collapsed Gibbs sampler.}
\end{figure}

\subsection*{Correlation of posterior sample}
\begin{figure}[H]
  \centering
  \includegraphics[
  width=\textwidth
  ]{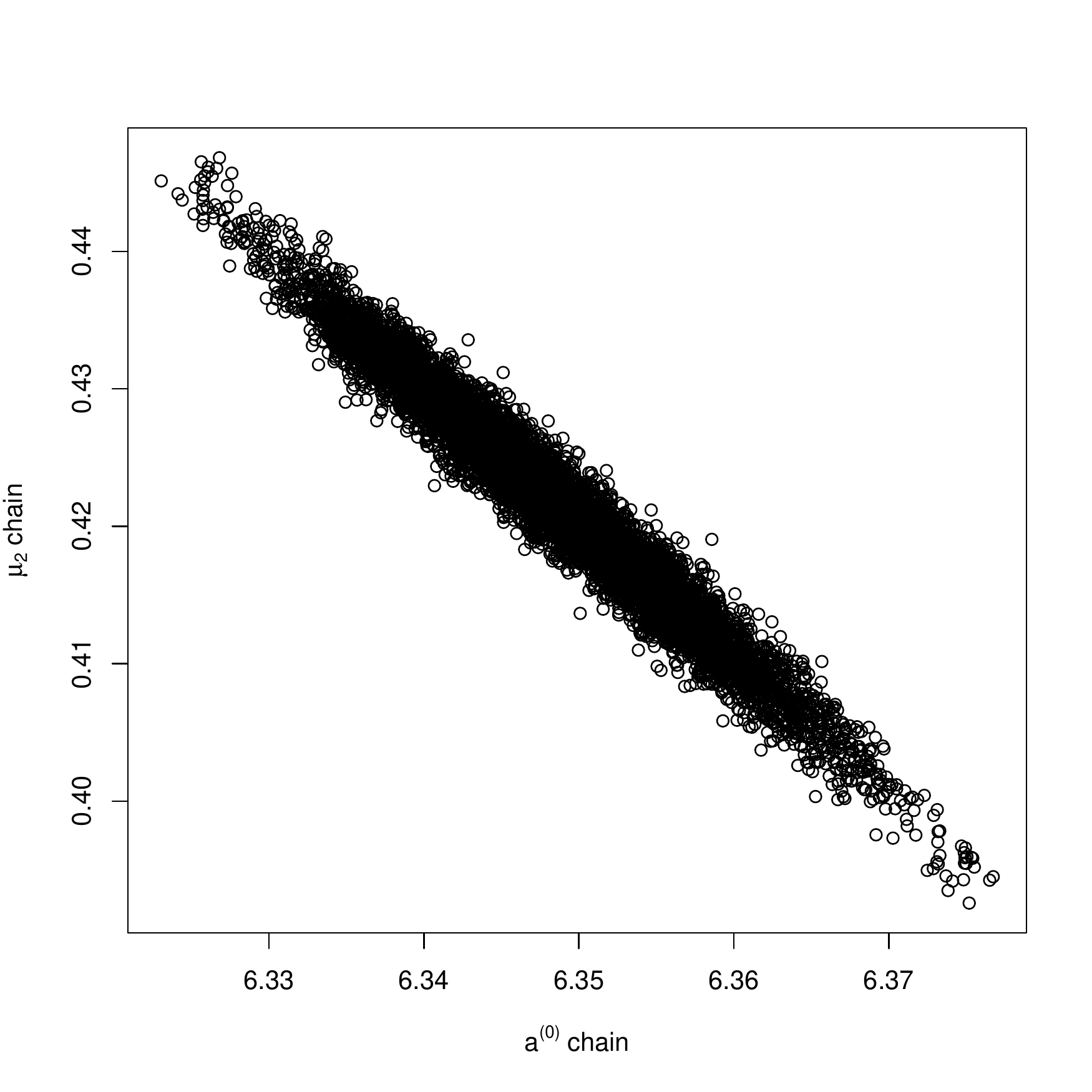}
 \caption{\label{fig:gibbsazeromu2}
Scatter plot of $\azero$ and $\mu_2$ chain.}
\end{figure}

Below we give the scatter plot of the posterior sample of $\azero$ and $\mu_2$ for Gibbs sampler on Stitch Fix data. Figure~\ref{fig:gibbsazeromu2} explains why we see a similar effective sample size for $\azero$ and $\mu_2$ chain.

\bibliographystyle{acm}
\bibliography{weaklybalancedreferencenew}

\begin{thebibliography}{10}

\bibitem{lme4}
{\sc Bates, D., M{\"a}chler, M., Bolker, B., and Walker, S.}
\newblock Fitting linear mixed-effects models using {lme4}.
\newblock {\em Journal of Statistical Software 67}, 1 (2015), 1--48.

\bibitem{gao2017efficient}
{\sc Gao, K., and Owen, A.}
\newblock Efficient moment calculations for variance components in large
  unbalanced crossed random effects models.
\newblock {\em Electronic Journal of Statistics 11}, 1 (2017), 1235--1296.

\bibitem{gao2016estimation}
{\sc Gao, K., and Owen, A.~B.}
\newblock Estimation and inference for very large linear mixed effects models.
\newblock {\em arXiv preprint arXiv:1610.08088\/} (2016).

\bibitem{gelman2005analysis}
{\sc Gelman, A.}
\newblock Analysis of variance—why it is more important than ever.
\newblock {\em The annals of statistics 33}, 1 (2005), 1--53.

\bibitem{gelmshir2011}
{\sc Gelman, A., and Shirley, K.}
\newblock {\em Inference from simulations and monitoring convergence}, vol.~6.
\newblock CRC Press Boca Raton, FL, 2011.

\bibitem{ghosh2020backfitting}
{\sc Ghosh, S., Hastie, T., and Owen, A.~B.}
\newblock Backfitting for large scale crossed random effects regressions.
\newblock {\em arXiv preprint arXiv:2007.10612\/} (2020).

\bibitem{ghosh2021scalable}
{\sc Ghosh, S., Hastie, T., and Owen, A.~B.}
\newblock Scalable logistic regression with crossed random effects.
\newblock {\em arXiv preprint arXiv:2105.13747\/} (2021).

\bibitem{Latala}
{\sc Lata{\l}a, R.}
\newblock Some estimates of norms of random matrices.
\newblock {\em Proceedings of the American Mathematical Society 133}, 5 (2005),
  1273--1282.

\bibitem{little2019statistical}
{\sc Little, R.~J., and Rubin, D.~B.}
\newblock {\em Statistical analysis with missing data}, vol.~793.
\newblock John Wiley \& Sons, 2019.

\bibitem{liu1994collapsed}
{\sc Liu, J.~S.}
\newblock The collapsed {Gibbs} sampler in {Bayesian} computations with
  applications to a gene regulation problem.
\newblock {\em Journal of the American Statistical Association 89}, 427 (1994),
  958--966.

\bibitem{papaspiliopoulos2020scalable}
{\sc Papaspiliopoulos, O., Roberts, G.~O., and Zanella, G.}
\newblock Scalable inference for crossed random effects models.
\newblock {\em Biometrika 107}, 1 (2020), 25--40.

\bibitem{papaspiliopoulos2021scalable}
{\sc Papaspiliopoulos, O., Stumpf-F{\'e}tizon, T., and Zanella, G.}
\newblock Scalable computation for {Bayesian} hierarchical models.
\newblock {\em arXiv preprint arXiv:2103.10875\/} (2021).

\bibitem{robertsahoo2001}
{\sc Roberts, G.~O., and Rosenthal, J.~S.}
\newblock Markov chains and de-initializing processes.
\newblock {\em Scandinavian Journal of Statistics 28}, 3 (2001), 489--504.

\bibitem{roberts1997updating}
{\sc Roberts, G.~O., and Sahu, S.~K.}
\newblock Updating schemes, correlation structure, blocking and
  parameterization for the {Gibbs} sampler.
\newblock {\em Journal of the Royal Statistical Society: Series B (Statistical
  Methodology) 59}, 2 (1997), 291--317.

\bibitem{sear:case:mccu:1992}
{\sc Searle, S.~R., Casella, G., and McCulloch, C.~E.}
\newblock {\em Variance Components}.
\newblock Wiley, New York, 1992.

\bibitem{vatsflegjone2019}
{\sc Vats, D., Flegal, J.~M., and Jones, G.~L.}
\newblock Multivariate output analysis for {Markov chain Monte Carlo}.
\newblock {\em Biometrika 106}, 2 (2019), 321--337.

\end{thebibliography}

\end{document}